\documentclass[12pt]{article}
\makeatletter \@addtoreset{equation}{section} \makeatother
\renewcommand{\theequation}{\thesection.\arabic{equation}}
\newcommand{\ba}{\begin{array}}
\newcommand{\ea}{\end{array}}
\newcommand{\beq}{\begin{equation}}
\newcommand{\eeq}{\end{equation}}
\newcommand{\bea}{\begin{eqnarray}}
\newcommand{\eea}{\end{eqnarray}}

\def\bce{\begin{center}}
\def\ece{\end{center}}

\def\nonu{\nonumber}

\def\pa{\partial}
\def\al{\alpha}
\def\be{\beta}

\def\de{\delta}

\def\ep{\epsilon}

\def\th{\theta}

\def\si{\sigma}

\newcommand{\dsqrt}[1]{\sqrt{\mathstrut #1}}

\def\eps6{{\displaystyle \mathop{\epsilon}^{6}}{}}
\def\g6{{\displaystyle \mathop{g}^{6}}{}}

\def\nab6{{\displaystyle \mathop{\nabla}^{6}}{}}

\def\0{{\sst{(0)}}}
\def\1{{\sst{(1)}}}
\def\2{{\sst{(2)}}}
\def\3{{\sst{(3)}}}
\def\4{{\sst{(4)}}}
\def\5{{\sst{(5)}}}
\def\6{{\sst{(6)}}}
\def\7{{\sst{(7)}}}
\def\8{{\sst{(8)}}}

\def\ba{\begin{array}}
\def\ea{\end{array}}
\def\beq{\begin{equation}}
\def\eeq{\end{equation}}
\def\be{\begin{equation}}
\def\ee{\end{equation}}

\def\eps{\epsilon}

\def\a{{\alpha}}

\def\th{{\theta}}

\def\ba{\begin{array}}
\def\ea{\end{array}}
\def\beq{\begin{equation}}
\def\eeq{\end{equation}}
\def\be{\begin{equation}}
\def\ee{\end{equation}}

\def\eps{\epsilon}

\def\a{{\alpha}}

\def\th{{\theta}}

\def\eps6{{\displaystyle \mathop{\epsilon}^{6}}{}}

\def\nab6{{\displaystyle \mathop{\nabla}^{6}}{}}

\newcommand{\bean}{\begin{eqnarray*}}
\newcommand{\eean}{\end{eqnarray*}}

\begin{document}
\thispagestyle{empty} \addtocounter{page}{-1}
   \begin{flushright}
KIAS-P10003 \\
\end{flushright}

\vspace*{1.3cm}
  
\centerline{ \Large \bf   Towards }
\vspace{.3cm} 
\centerline{ \Large \bf   An ${\cal N}=1$ 
$SU(3)$-Invariant Supersymmetric
  Membrane Flow }
\vspace{.3cm} 
\centerline{ \Large \bf In Eleven-Dimensional Supergravity }
\vspace*{1.5cm}
\centerline{{\bf Changhyun Ahn 
 {\rm and} Kyungsung Woo }
} 
\vspace*{1.0cm} 
\centerline{\it  
Department of Physics, Kyungpook National University, Taegu
702-701, Korea} 
\vspace*{0.8cm} 
\centerline{\tt ahn@knu.ac.kr 
\qquad wooks@knu.ac.kr
} 
\vskip2cm

\centerline{\bf Abstract}
\vspace*{0.5cm}

The M-theory lift of ${\cal N}=1$ 
$G_2$-invariant RG flow via a combinatoric use of  
the 4-dimensional RG flow and 11-dimensional Einstein-Maxwell
equations
was found some time ago. 
The 11-dimensional metric, a warped product of an asymptotically
$AdS_4$ space with a squashed and stretched 7-sphere, for $SU(3)$-invariance 
was found before.
In this paper,
by choosing the 4-dimensional internal space as 
${\bf CP}^2$ space, 
we discover an  exact solution of ${\cal N}=1$ $G_2$-invariant flow 
to the 11-dimensional field equations. 
By an appropriate coordinate transformation on the three
internal coordinates, 
we also find an  11-dimensional solution 
of ${\cal N}=1$ $G_2$-invariant flow interpolating from ${\cal N}=8$
$SO(8)$-invariant UV fixed point to ${\cal N}=1$ $G_2$-invariant IR
fixed point. In particular, the 11-dimensional metric and 4-forms at
the ${\cal N}=1$ $G_2$ fixed point for the second solution will
provide some hints for the 11-dimensional lift of 
whole ${\cal N}=1$ $SU(3)$ RG flow
connecting  this ${\cal N}=1$ $G_2$ fixed point 
to ${\cal N}=2$ $SU(3) \times U(1)_R$ fixed point in 4-dimensions.

\baselineskip=18pt
\newpage
\renewcommand{\theequation}
{\arabic{section}\mbox{.}\arabic{equation}}

\section{Introduction}

The low energy limit of $N$ membranes at 
${\bf C}^4/{\bf Z}_k$ singularity is described
in the context of 3-dimensional 
${\cal N}=6$ $U(N) \times U(N)$ 
Chern-Simons matter theory
with level $k$ \cite{ABJM}.
For particular level $k=1, 2$, the maximal ${\cal N}=8$ supersymmetry
is preserved. 
The matter contents and the superpotential  of this theory
coincide with  the ones in the theory for 
D3-branes at the conifold in 4-dimensions \cite{KW}. 
The renormalization group (RG) flow of the 3-dimensional 
theory can be studied from the 4-dimensional ${\cal N}=8$ gauged
supergravity  via $AdS_4$/$CFT_3$ 
correspondence \cite{Maldacena}. 
The holographic ${\cal N}=2$ supersymmetric
$SU(3) \times U(1)_R$-invariant RG flow 
connecting the maximally supersymmetric ${\cal N}=8$ $SO(8)$
ultraviolet (UV) point 
to ${\cal N}=2$ $SU(3) \times U(1)_R$ infrared (IR) point has been found in 
\cite{AP,AW,AR99} while
the ${\cal N}=1$ supersymmetric $G_2$-invariant RG flow  
from this maximally supersymmetric  ${\cal N}=8$ $SO(8)$  UV point 
to
${\cal N}=1$ $G_2$ IR point has been discussed in 
\cite{AW,AI}.
The former has $SU(3) \times U(1)_R$ symmetry around IR region
including the IR critical point
and the latter has $G_2$ symmetry around IR region as well as the IR
critical point.
The 11-dimensional M-theory lifts of these two RG flows 
have been described in \cite{CPW,AI}  by solving the
Einstein-Maxwell equations explicitly in 11-dimensions.

The mass deformed $U(2) \times U(2)$
Chern-Simons matter theory with level $k=1, 2$ 
preserving the above global ${\cal N}=2$ 
$SU(3) \times U(1)_R$ symmetry has been found 
in \cite{Ahn0806n2,BKKS} while
the mass deformation for this theory preserving ${\cal N}=1$ $G_2$
symmetry  has been found in \cite{Ahn0806n1}.  
Further nonsupersymmetric 
RG flow equations preserving two $SO(7)^{\pm}$ symmetries 
have been studied in \cite{Ahn0812}.  
The holographic ${\cal N}=1$ supersymmetric $SU(3)$-invariant
RG flow equations connecting  ${\cal N}=1$ $G_2$ point 
to  ${\cal N}=2$ $SU(3) \times U(1)_R$ point in 4-dimensions have been 
studied in \cite{BHPW}.  
Moreover, the other holographic supersymmetric
RG flows have been 
found and  
further developments on  
the 4-dimensional gauged supergravity (see also \cite{GW02,PW03}) 
have been made
in \cite{AW09,Ahn0905}. 

The spin-2 Kaluza-Klein modes around 
a warped product of $AdS_4$ and a seven-ellipsoid having 
above global ${\cal N}=1$ $G_2$ symmetry are discussed in \cite{AW0907}. 
The gauge dual with 
${\cal N}=2$ $SU(2) \times SU(2) \times U(1)_R$ symmetry 
for the 11-dimensional lift of 
$SU(3) \times U(1)_R$-invariant solution 
in 4-dimensional supergravity 
is described in \cite{AW0908} (see also \cite{FKR}). 
The 11-dimensional description preserving ${\cal N}=2$ $SU(2) \times U(1)
\times U(1)_R$ symmetry is found in \cite{Ahn0909} and
the smaller ${\cal N}=2$ $U(1) \times U(1) \times U(1)_R$ symmetry flow is
discussed in \cite{Ahn0910}.
All of these have common $U(1)_R$ factor corresponding to ${\cal N}=2$
supersymmetry. The symmetries come from the symmetry each 
5-dimensional Sasaki-Einstein space has.    

When
the 11-dimensional theory from the 4-dimensional gauged ${\cal N}=8$
supergravity is constructed, the various
11-dimensional solutions  occur even though the flow equations
characterized by the 4-dimensional supergravity fields 
are
the same. This is mainly due to the fact that 
the 11-dimensional metric with common geometric parameters 
is different from each other.
The same flow equations \cite{AP,AI,AW} in 4-dimensions, playing the
role of geometric parameters in the internal space, provide
different 11-dimensional solutions to the equations of the
motion for 11-dimensional supergravity \cite{CPW,AW0908,Ahn0909,Ahn0910}.
The invariance of 11-dimensional metric and 4-forms determines 
the global symmetry. Sometimes the 4-forms restrict to the global
symmetry the metric possesses and break further into the smaller symmetry group. 

The compact 7-dimensional manifold in the compactification of
11-dimensional supergravity can be described by the metric 
encoded in the vacuum expectation values for 4-dimensional ${\cal
N}=8$ gauged supergravity \cite{dN}. 
The $SU(3)$-singlet space contains various
critical points \cite{Warner83} and 
RG flows (domain walls) along the $AdS_4$ radial coordinate.   
Based on the nonlinear metric ansatz by \cite{dWNW}, the geometric construction of
the compact 7-dimensional metric is found and the local frames  
are obtained by decoding the $SU(3)$-singlet vacuum expectation values
into squashing and stretching parameters of the 7-dimensional manifold
\cite{AI02}.  
The M-theory lift of a supersymmetric RG flow can be done, in general, as follows. 
We impose the nontrivial $AdS_4$ radial coordinate dependence of vacuum expectation
values subject to the 4-dimensional RG flow equations \cite{AP,AI,AW}. Then the
geometric parameters in the 7-dimensional metric at certain critical
point are controlled by the RG flow equations so that they can be
extrapolated from the critical points. Next,  we make an appropriate ansatz 
for the 11-dimensional 4-form field strengths. 
Finally, the 11-dimensional Einstein-Maxwell bosonic equations \cite{CJS,DNP} 
can be solved by using the RG flow equations \cite{AP,AI,AW} to complete the M-theory
uplift.

The $SU(3)$-singlet space with a breaking of the $SO(8)$ gauge group
into a group which contains $SU(3)$ can be represented by four real
supergravity fields \cite{Warner83}.
Although the final goal is to solve the 11-dimensional
Einstein-Maxwell equations to complete the 11-dimensional lift of the
whole $SU(3)$-invariant sector (that contains four supergravity
fields) 
including the RG flows \cite{AW}, 
we focus on the $G_2$-invariant sector (which is a subset of above
$SU(3)$-invariant sector) in 4-dimensional viewpoint by constraining
the four arbitrary supergravity fields together with the particular condition.
There exist a supersymmetric ${\cal N}=1$ $G_2$ critical point and two
nonsupersymmetric $SO(7)^{\pm}$ critical points \cite{dN84,Englert} in this sector. 
If we take the different constraints on the supergravity fields, 
then we are led to the $SU(3) \times U(1)_R$-invariant sector where 
there are a supersymmetric ${\cal N}=2$ $SU(3) \times U(1)_R$ critical point and a
nonsupersymmetric $SU(4)^{-}$ critical point \cite{PW}.  

The 11-dimensional metric is found by \cite{AI02} where the compact
7-dimensional metric and warp factor are completely determined in
local frames. The two geometric parameters, by constraints, 
that are nothing but the
above supergravity fields, depend on the $AdS_4$
radial coordinate and are subject to the RG flow equations \cite{AI,AW} in
4-dimensional gauged supergravity.
The global coordinates for ${\bf S}^7$ appropriate for the base
six-sphere ${\bf S}^6 \simeq \frac{G_2}{SU(3)}$ preserve the
Fubini-Study metric on ${\bf CP}^2$ and this describes the
ellipsoidally deformed ${\bf S}^7$ \cite{dWNW}.
On the other hand, 
by using the relation to the Hopf fibration on ${\bf CP}^3 \simeq
\frac{SU(4)}{SU(3) \times U(1)}$ explicitly
and
keeping only the ${\bf CP}^2$ space, one can
change the remaining three local frames in 7-dimensional manifold via an
orthogonal transformation \cite{AI02}.
Each global coordinates depends on each base six-spheres they use. 

What is the 11-dimensional lift of holographic ${\cal N}=1$
supersymmetric $SU(3)$-invariant RG flow \cite{BHPW} 
connecting from ${\cal N}=1$ $G_2$ critical point to 
${\cal N}=2$ $SU(3) \times U(1)_R$ critical
point? 
At each critical point, the 4-forms are known in different
background. 
Along the supersymmetric RG flow, one 
expects that there exist the extra 4-forms which should vanish at the
critical points. 
In order to describe the whole RG flow, one needs to have the
consistent background. Around ${\cal N}=2$ $SU(3) \times U(1)_R$ IR fixed point, 
the use of global coordinates for Hopf fibration on ${\bf CP}^3$ was
done in \cite{CPW}. Around ${\cal N}=1$ $G_2$ IR fixed point, 
the global coordinates for ${\bf S}^7$ appropriate for the base
six-sphere ${\bf S}^6$ was used in \cite{AI}. Therefore, 
either one should find the ${\bf CP}^3$-basis around ${\cal
  N}=1$ IR fixed point, or one needs to find ${\bf S}^6$-basis
around
${\cal N}=2$ IR fixed point. Recently, the latter is studied in \cite{AW1004}. 
Now then we are left with the former.

In this paper, 
we find out an  exact solution of ${\cal N}=1$ $G_2$-invariant flow 
(connecting from the ${\cal N}=8$ $SO(8)$ UV fixed point to 
${\cal N}=1$ $G_2$ IR fixed point) 
to the 11-dimensional Einstein-Maxwell
equations where the internal space contains the ${\bf CP}^2$ space. 
By an appropriate coordinate transformation, we also find a 
 11-dimensional solution 
of ${\cal N}=1$ $G_2$-invariant flow 
(connecting from the ${\cal N}=8$ $SO(8)$ UV fixed point to 
${\cal N}=1$ $G_2$ IR fixed point) in different background where
the rectangular coordinates we use are the same as the ones in \cite{CPW}. 
The two solutions share the same geometric
parameters
satisfying the common RG flow equations \cite{AI} in 4-dimensional 
gauged ${\cal N}=8$ supergravity.
At the ${\cal N}=1$ IR critical point, we have found both the metric
and 4-forms explicitly. They are completely different from the
previous findings in \cite{AI} in the sense that the rectangular
coordinates with specific angular coordinates are the same as the one
in \cite{CPW} and they are different parametrization from the one in \cite{AI}.
This finding will give us some hints for the final goal,
the 11-dimensional lift of ${\cal N}=1$
supersymmetric $SU(3)$-invariant RG flow  
connecting from ${\cal N}=1$ $G_2$ fixed point to 
${\cal N}=2$ $SU(3) \times U(1)_R$ fixed point.

In section 2, we introduce
the six basis vectors  living in the unit round six-sphere ${\bf S}^6$ 
and the normal vector perpendicular to ${\bf S}^6$ 
in the context of octonion  description \cite{GW}.
The three $G_2$-invariant tensors are determined by the components of
these seven frames.
Then the 4-forms are given by these $G_2$-invariant tensors as well as a
geometric superpotential \cite{AI}. 
By applying these 4-forms with varying scalars, the exact solution
\cite{AI} to
the 11-dimensional Einstein-Maxwell equations corresponding to the
11-dimensional lift of the ${\cal N}=1$ $G_2$-invariant RG flow is reviewed. 

By taking the different spherical parametrization where the ${\bf
CP}^2$
structure is evident,
the six basis vectors  living in the six-sphere ${\bf S}^6$ 
and the normal vector perpendicular to it are constructed explicitly.
The three $G_2$-invariant tensors are also determined by the components of
these new seven frames.
Assuming that the two supergravity fields satisfy the domain
wall solutions \cite{AI} as before, 
we compute the Ricci tensor in this background completely.  
Surprisingly, the Ricci tensor  has the 
same expression as the one in previous  spherical parametrization.
This indicates that the 4-form field strengths are constructed
similarly by using the above new three $G_2$-invariant tensors and a
geometric superpotential introduced in \cite{AI}. 
Eventually, we determine the  solution for 
the 11-dimensional Einstein-Maxwell equations  corresponding to the
the different 11-dimensional lift of the same ${\cal N}=1$ $G_2$-invariant RG flow.
The 4-forms for both parametrizations 
depend on the 7-dimensional internal coordinates via each invariant tensors. 

In section 3, by changing only three of them among seven
internal coordinates characterized by previous second spherical
parametrization, the 11-dimensional metric can be written in terms of these new
coordinates which preserve the Fubini-Study metric on ${\bf CP}^2$ \cite{AI02}.
The Ricci tensor can be expressed as a linear combination of the Ricci
tensor for $G_2$-invariant case and similarly the 4-forms also are given by 
a linear combination of 4-forms for $G_2$-invariant flow.  
Then, we find out a  solution for 
the 11-dimensional Einstein-Maxwell equations  corresponding to the
11-dimensional lift of the ${\cal N}=1$ $G_2$-invariant RG flow
connecting from the ${\cal N}=8$ $SO(8)$ UV fixed point to 
${\cal N}=1$ $G_2$ IR fixed point.
Since we focus on the $G_2$-invariant sector, the two geometric
parameters satisfy the RG flow equations \cite{AI}. In the 11-dimensional point
of view, both the metric and 4-forms preserve the $G_2$ symmetry. 

In section 4, we summarize the results of this paper and present some
future directions.

In the Appendices, we present the detailed expressions for the 
Ricci tensor, 4-form field strengths, seven frames, three invariant tensors and
Maxwell equations.

\section{An ${\cal N}=1$ $G_2$-invariant 
supersymmetric flow in an 11-dimensional theory}

We construct three $G_2$-invariant tensors in terms of six-sphere
coordinates explicitly in the subsection 2.1 and 
find out a  ${\cal N}=1$ $G_2$-invariant  solution for 
11-dimensional Einstein-Maxwell equations in subsection 2.2. 

\subsection{Spherical parametrization I}

Let us consider the spherical parametrization describing the unit
round five-sphere ${\bf S}^5$
\bea
u^1 & = & \sin \theta_1 \, \sin \theta_2 \, \sin \theta_3 \, \sin \theta_4
\, \sin \theta_5,
\nonu \\
u^2 & = & \cos \theta_1 \, \sin \theta_2 \, \sin \theta_3 \, \sin \theta_4
\, \sin \theta_5,
\nonu \\
u^3 & = & \cos \theta_2 \, \sin \theta_3 \, \sin \theta_4
\, \sin \theta_5,
\nonu \\
u^4 & = & \cos \theta_3 \, \sin \theta_4
\, \sin \theta_5,
\nonu \\
u^5 & = & \cos \theta_4
\, \sin \theta_5,
\nonu \\
u^6 & = & \cos \theta_5,
\label{sixu}
\eea
subject to the constraint $\sum_{i=1}^{6} (u^i)^2=1$.
In order to use the 6-dimensional coordinatization for unit round 
six-sphere ${\bf S}^6$, 
let us also introduce the two orthogonal unit vectors
in ${\bf R}^7$ with (\ref{sixu}) 
\bea
{\bf U} =(u^1, u^2, u^3, u^4, u^5, u^6, 0), \qquad 
{\bf V} =(0, 0, 0, 0, 0, 0, 1).
\label{UV}
\eea
Then the unit normal vector perpendicular 
to the six-sphere ${\bf S}^6$ can be identified with
\bea
\hat{\bf n} = {\bf U} \, \sin \theta_6 + {\bf V} \, \cos \theta_6.
\label{normal}
\eea
The unit magnitude of normal vector, $(\hat{\bf n}, \hat{\bf n})=1$, from 
(\ref{UV}) can be easily checked.
The rectangular coordinates parametrizing the six-sphere ${\bf S}^6$ inside ${\bf
R}^7$ are given by $X^i = u^i \, \sin \theta_6(i =1, \cdots, 6)$ 
and $X^7=\cos
\theta_6$ with $\sum_{A=1}^{7} (X^A)^2 =1$ whose gradient is exactly the unit
normal vector introduced in (\ref{normal}). This parametrization for 
$X^A$ can be also obtained
from the one corresponding to the unit round seven-sphere ${\bf S}^7$
by ignoring seventh coordinate $\theta \equiv y^7$ in \cite{AI}. 
By taking the gradient to the normal vector with respect to the six
internal coordinates, 
a set of basis vectors spanning a tangent space on the six-sphere ${\bf S}^6$ is
given by \cite{GW}
\bea
\hat{\bf e}_i = - \frac{\pa \hat{\bf n}}{\pa \theta_i}, \qquad i=1,
\cdots, 6,
\label{ei}
\eea
which is orthogonal to the normal vector (\ref{normal}).
In general, the second fundamental tensor arises in (\ref{ei}) for
generic hypersurface. For round six-sphere, this second fundamental
tensor is proportional to the 6-dimensional metric for six-sphere and
this leads to a simple relation (\ref{ei}) above. The orthonormality
for the basis vectors (\ref{ei}) can be checked also. 

Adding the unit normal vector $\hat{\bf n}$ as the seventh-frame to the six-sphere 
${\bf S}^6$-frames $\hat{\bf e}_i$ provides the tangent frames of 
Cayley space denoted by ${\bf I}^7$ \cite{GW}.
\bea
\hat{\bf e}_i = \hat{\bf e}_i^A \, O_A, \qquad
\hat{\bf n} = \hat{\bf n}^A \, O_A, \qquad A=1, \cdots, 7,
\label{eio}
\eea
where $O_A$ are imaginary octonions satisfying the algebra
\bea
O_A O_B = -\delta_{AB} + \eta_{ABC} \, O_C,
\label{algebra}
\eea
with completely antisymmetric structure constants $\eta_{ABC}$ \cite{GG}.
The nonzero $\eta_{ABC}$  are given by 1 for the indices 
$ABC=123, 516, 624, 435, 471,
673, 572$.
In order to obtain the $G_2$-invariant tensors on ${\bf S}^6$,
one projects ${\bf I}^7$ invariant tensor and its dual to local
tangent frames and obtains 7-dimensional tensor (and its dual).
Since the geometry of the $G_2$-invariant ${\bf S}^6$
is already built in the 7-dimensional coordinatization,
what we have to do is to take the dimensional reduction of
7-dimensional tensor (and its dual) to the 6-dimensional ones. 
Then the almost complex structure $F_{ij}$, its covariant derivative
$T_{ijk}$, and its Hodge dual, $S_{ijk}$ that depend on 
the internal coordinates $\theta_i(i=1, 2, \cdots, 6)$, 
with (\ref{normal}), (\ref{ei}) and (\ref{eio}), can be
summarized by 
\bea
F_{ij} & = & -\eta_{ABC} \, \hat{\bf e}^A_i  \, \hat{\bf e}^B_j  \, \hat{\bf n}^C, 
\nonu \\
T_{ijk} & = & \eta_{ABC} \, \hat{\bf e}^A_i  \, \hat{\bf e}^B_j  \, \hat{\bf e}^C_k, 
\nonu \\
S_{ijk} & = & \eta_{ABCD} \, \hat{\bf e}^A_i  \, \hat{\bf e}^B_j  \, 
\hat{\bf e}^C_k \, \hat{\bf n}^D,
\label{fts}
\eea
where the dual tensor \cite{WJ} $\eta_{ABCD}$ of the structure
constant 
$\eta_{EFG}$ is given by 
$\eta_{ABCD} = \eta_{ABE} \eta_{CDE} -\delta_{AC} \delta_{BD} + 
\delta_{AD} \delta_{BC}$ explicitly. The nonzero components for dual
tensor can be
read off from this relation. Note that $\eta_{ABC}$ is ${\bf
  I}^7$-invariant tensor while $\eta_{DEFG}$ is its dual.
The dimensional reduction of 7-dimensional tensor leads to 
$F_{ij}$ and $T_{ijk}$ tensors while 
the dimensional reduction of its dual provides 
$S_{ijk}$ tensor, as shown in (\ref{fts}).  
The last relation of (\ref{fts}) can be 
checked from the definition of $S_{ijk} = F_i^{\, l} T_{ljk}$ \cite{GW} by using
the first two relations of (\ref{fts}) together with (\ref{algebra}). 
We can 
introduce the dot product and cross product for arbitrary imaginary
octonions \cite{GW}
and these can be used for the construction of 7-dimensional tensor and its dual.

By computing the dot product between $\hat{\bf e}_i$ and $\hat{\bf e}_j$
in ${\bf I}^7$, one obtains the 6-dimensional metric on ${\bf S}^6$
for the spherical parametrization we start with.
That is, as we expect, 
\bea
g_{ij} = \mbox{diag} (s_2^2 \, s_3^2 \, s_4^2 \, s_5^2 \, s_6^2, 
\,\,\, s_3^2 \, s_4^2 \, s_5^2 \, s_6^2, \,\,\, s_4^2 \, s_5^2 \, s_6^2,  \,\,\, 
s_5^2 \, s_6^2, \,\,\, s_6^2, \,\,\, 1), \qquad 
s_i \equiv \sin \th_i.
\label{6d}
\eea
Now it is ready to construct the 7-dimensional ellipsoidal metric.
Applying the Killing vector to the metric formula \cite{dWNW}, one
obtains the 7-dimensional inverse metric with undetermined warp
factor.
Then plugging this inverse metric into the definition of warp factor, 
the warp factor is fixed. Finally, substituting the warp factor into
the inverse metric leads to our 7-dimensional metric. 
The 7-dimensional warped ellipsoidal metric 
used in \cite{dWNW,AI} can be summarized as
\bea
d s_7^2  &=&   \sqrt{\Delta \, a} \, L^2 \left[ \frac{\xi^2}{a^2} \, d
  \theta^2 +
\sin^2 \theta  \, d \Omega_6^2 \right], \qquad
d \Omega_6^2 =  d \theta_6^2 + \sin^2 \theta_6 \, d \Omega_5^2, 
\nonu \\
d \Omega_5^2 & = & d \theta_5^2 + \sin^2 \theta_5\, \left(d \theta_4^2 +
\sin^2 \theta_4\, \left [ d \theta_3^2 +
\sin^2 \theta_3 \, \left( d \theta_2^2 +
\sin^2 \theta_2\, d \theta_1^2 \right) \right] \right), 
\label{met1}
\eea
where the quadratic form $\xi^2$ is given by
\bea
\xi^2 = a^2 \, \cos^2 \theta + b^2 \, \sin^2 \theta,
\label{xi2}
\eea
and the warp factor together with (\ref{xi2}) is also given by
\bea
\Delta = a^{-1} \, \xi^{-\frac{4}{3}}.
\label{warp}
\eea
The standard 7-dimensional ellipsoidal metric is warped by 
a factor $\sqrt{\Delta \, a}$ in (\ref{met1}). The 6-dimensional
metric $d \Omega_6^2$ is equal to (\ref{6d}). 
The two vacuum expectation values $(a, b)$ in 4-dimensional gauged
${\cal N}=8$ supergravity that appear in the 7-dimensional internal 
metric (\ref{met1}) are functions of the $AdS_4$ radial
coordinate $r \equiv x^4$.
The eccentricity for 7-dimensional ellipsoid is given by $\sqrt{1-\frac{b^2}{a^2}}$.

The 11-dimensional metric \cite{dWNW,dN87} 
by combining the above 7-dimensional metric
(\ref{met1})
with the 4-dimensional metric together with warp factor (\ref{warp}) yields
\bea
d s_{11}^2 = \Delta^{-1} \, \left( d r^2 + e^{2A(r)}\, \eta_{\mu \nu}
  \, d x^{\mu} \, d x^{\nu} \right) + d s_7^2,
\label{11met}
\eea
where $r \equiv x^4$ and $\mu, \nu=1, 2, 3$ with $\eta_{\mu \nu}
=\mbox{diag}(-, +, +)$.   
The 11-dimensional coordinates
with indices $M, N, \cdots$ are decomposed into 4-dimensional spacetime 
$x^{\mu}$ and
7-dimensional internal space $y^m$.
Denoting the 11-dimensional metric as $g_{MN}$ with 
the convention 
$(-, +, \cdots, +)$
and the antisymmetric 
tensor fields as $F_{MNPQ}$, the 11-dimensional
Einstein-Maxwell equations are given by \cite{CJS,DNP}
\bea
R_{M}^{\;\;\;N} & = & \frac{1}{3} \,F_{MPQR} F^{NPQR}
-\frac{1}{36} \de^{N}_{M} \,F_{PQRS} F^{PQRS},
\nonu \\
\nabla_M F^{MNPQ} & = & -\frac{1}{576} \,E \,\ep^{NPQRSTUVWXY}
F_{RSTU} F_{VWXY},
\label{fieldequations}
\eea
where the covariant derivative $\nabla_M$ 
on $F^{MNPQ}$ in 
(\ref{fieldequations})
is given by 
$E^{-1} \pa_M ( E F^{MNPQ} )$ together with elfbein determinant 
$E \equiv \sqrt{-g_{11}}$. The epsilon tensor 
 $\ep_{NPQRSTUVWXY}$ with lower indices is purely numerical.
For given 11-dimensional metric, the nontrivial task is to find the
correct 4-forms that satisfy (\ref{fieldequations}).

The warped 11-dimensional metric (\ref{11met}) with (\ref{met1}) generates the Ricci
tensor \cite{AI} that depends on  $(a, b, A)$ and their derivatives
with respect to $r$ and $\theta$. Applying the flow equations, 
all the $r$-derivatives in the Ricci tensor can be replaced with the
functions of $(a, b)$.  
The Ricci tensor is given in the Appendix $A$ explicitly 
after substituting the flow equations \cite{AI,AW}
\bea
\frac{d  a}{d r} &=& -\frac{8}{7L}
\biggl[a^2\,\partial_a W +(ab-2)\,\partial_{b} W \biggr] 
\nonu \\
& = & -\frac{a^{\frac{3}{2}} \left[ a^5 - 88 \, a^2 \, b + 14 \, a^3  \,
      b^2 -56 \, b^3 + a (80 + 49 \, b^4) \right]}{2\, L\,
\sqrt{(a^2 +7b^2)^2 -112\,(ab-1)} },
\nonu \\
\frac{d b}{d r} &=& -\frac{8}{7L}
\biggl[b^2\,\partial_{b} W +(ab-2)\,\partial_a W \biggr]
\nonu \\
& = & \frac{\sqrt{a} \, \left[ 96 + 2\, a^4 - a (176+ a^4) \, b + 100
    \, a^2 \, b^2 - 14 \, a^3 \, b^3 + 42 \, b^4 - 49 \, a \, b^5
\right]}{2 \, L\, \sqrt{(a^2 +7b^2)^2 -112\,(ab-1)}},
\nonu \\
\frac{d A}{d r} &=& \frac{2}{L} \, W = \frac{1}{4\, L} \, a^{3 \over
  2} \,  \sqrt{(a^2 +7b^2)^2 -112\,(ab-1)},
\label{flow}
\eea
where the superpotential $W$ in 4-dimensional gauged ${\cal N}=8$ supergravity 
is given by a function of $(a,b)$:
\begin{equation}
W= {1 \over 8}\,a^{3 \over 2}
\dsqrt{(a^2 +7b^2)^2 -112\,(ab-1)}.
\label{spoten}
\end{equation}
There exist only two nonzero off-diagonal components $R_4^{\,\,5}$ and
$R_5^{\,\,4}$
as well as nonzero diagonal components.
In 4-dimensions, there exist two critical points, ${\cal N}=8$ $SO(8)$ critical
point at which $(a, b)=(1,1)$ and ${\cal N}=1$
$G_2$ critical point at which $(a, b)=(\sqrt{\frac{6\sqrt{3}}{5}}, 
\sqrt{\frac{2\sqrt{3}}{5}})$. 
At these two points, 
$\frac{d a}{d r}$
and $\frac{d b}{d r}$ vanish due to the right hand sides of 
(\ref{flow}) are equal to zero.
The criticality can be observed from the fact that the
first two right hand sides of (\ref{flow}) are written as the
derivatives of superpotential $W(a, b)$ 
with respect to the field $a$ and the
field $b$. 
The superpotential has $1$ and $\sqrt{\frac{36\sqrt{2} 
\, 3^{\frac{1}{4}}}{25\sqrt{5}}}$ at two
critical values respectively.

What about 4-form field strengths for 11-dimensional solution?
By interpreting the 3-form gauge field with membrane indices
as a geometric superpotential, which is a generalization of
4-dimensional superpotential (\ref{spoten}), times 
volume form and putting the previous $G_2$-invariant tensors (\ref{fts}) 
to various 3-form components, one can construct 
the most general $G_2$-invariant ansatz.    
The field strengths are summarized by \cite{AI}
\begin{eqnarray}
F_{\mu\nu\rho 4} & = & e^{3A(r)}
\,W_r (r,\theta)\,\epsilon_{\mu\nu\rho},\qquad
F_{\mu\nu\rho 5} = e^{3A(r)}
\,W_{\theta}(r,\theta)\,\epsilon_{\mu\nu\rho},
\nonu \\
F_{mnpq} &=& 2\,h_2 (r,\theta)\,\eps6_{mnpqrs} \, F^{rs}, \qquad
F_{5mnp} = \tilde{h}_1 (r,\theta)\,T_{mnp}
+\tilde{h}_2 (r,\theta)\,S_{mnp}, \nonumber\\
F_{4mnp} &=& \tilde{h}_3 (r,\theta)\,T_{mnp}
+\tilde{h}_4 (r,\theta)\,S_{mnp}, \qquad
F_{45mn} = \tilde{h}_5 (r,\theta)\,F_{mn}, 
\label{fst1}
\end{eqnarray}
where the eight 
coefficient functions $(W_r, W_{\theta}, \tilde{h}_1, h_2, \tilde{h}_2,
\tilde{h}_3,
\tilde{h}_4, \tilde{h}_5)$ which depend on both $r$ and $\theta$
are given by \cite{AI} and also they are in the Appendix $B$ for convenience.
The fields $(W_r, W_{\theta})$ are related to the geometric
superpotential $\widetilde{W}(r,\theta)$ which corresponds to the 3-form gauge
field with membrane indices, as mentioned before.
Compared with the previous works \cite{dWNW,GW} which holds for the
critical points only, the mixed field strengths $F_{1235}, F_{4mnp}$ and
$F_{45mn}$ were new \footnote{One can write down the 2-form and
  3-forms in terms of rectangular coordinates $X^i=u^i \, \sin \th_6
  \sin \th$($i=1, \cdots, 6$), 
$X^7=\cos \th_6 \, \sin \th$ and $X^8=\cos \th$. Since one can express the
angles in terms of $X^A$'s as $\cos \th_i =
\frac{X^{i+1}}{\sqrt{\sum_{j=1}^{i+1} (X^j)^2}}$ where $i=1, \cdots,
6$ and $\cos \th = X^8$, 
the rank 2 and
rank 3 tensors with angular coordinates can be written in terms of those with
$X^A$'s.  Let us write the 2-form $F_{mn} \, d \th^m \wedge d \th^n$ by
using the relation between $(\th, \th_i)$ and $X^A$ and changing 
the differentials $d \th^m$ in terms of $d X^A$'s.
Then it turns out that the 2-form is given by $F^{(2)} = \frac{
1 }{(\sum_{A=1}^7 (X^A)^2)^{\frac{3}{2}}} \, \eta_{ABC} \, X^A d X^B \wedge d X^c$ where
$\eta_{ABC}$ is the same as the one in (\ref{fts}) and the $X^8$ is $G_2$-singlet.  
The $G_2$-invariant 
structure constant $\eta_{ABC}$ is contracted  with the ${\bf 7}$'s 
of $G_2$ in $X^A$($A=1, \cdots, 7$) in this way.
Similarly the 3-form $S_{mnp} \, d \th^m \wedge d \th^n \wedge d \th^p$ 
is given by $S^{(3)} = \frac{
1 }{(\sum_{A=1}^7 (X^A)^2)^2} \, \eta_{ABCD} \, X^A d X^B \wedge d X^c \wedge d X^D$
where $\eta_{ABCD}$ is the same as the one in (\ref{fts})
and
the 3-form  
$T^{(3)} = d F^{(2)}$ is given. Then one arrives at 
$F^{(4)} = d A^{(3)} + h_2 \, d S^{(3)}+
d r \wedge (\tilde{h}_3 \, d F^{(2)} + \tilde{h}_4 \, S^{(3)})
+\frac{1}{\sqrt{1-(X^8)^2}}
d X^8 \wedge (\tilde{h}_1 \, d F^{(2)} + \tilde{h}_2 \, S^{(3)}) -
\frac{\tilde{h}_5}{\sqrt{1-(X^8)^2}} d r \wedge F^{(2)} \wedge d
X^8$ where the 3-form $A^{(3)}= - e^{3A } \, \widetilde{W} \, d x^1 \wedge d
x^2 \wedge d x^3$.}. 

The details of Einstein equations, the first equation of
(\ref{fieldequations}), 
are given in \cite{AI} explicitly.
The nonzero components of the Maxwell equation, the second equation of
(\ref{fieldequations}), are characterized by the following free indices 
\bea
(123), \qquad (4np), \qquad (5np), \qquad (mnp), \qquad m,n,p=6,
\cdots, 11, 
\label{max}
\eea
with the number of components $1$, $15$ by choosing two out of six, 
$15$ by choosing two out of six and $20$ by choosing three out of six respectively. 
Other remaining components of the Maxwell equation become identically zero. 
Therefore, there exist 51 nonzero components of the Maxwell equation. 
The right hand side of Maxwell equations for the $(123)$-component
consists of 35 terms coming from the quadratic 4-forms.
For $(4np)$- and $(5np)$-components, the right hand side of Maxwell
equations contains
only a single term in quadratic 4-forms. The former has $F_{1235}$ while
the latter has $F_{1234}$.
For the $(mnp)$-components, the two contributions from the quadratic
4-forms arise in the right hand side of Maxwell equations.
In this case, both 4-forms $F_{1235}$ and $F_{1234}$ appear.
The elfbein determinant $E=\sqrt{-g_{11}}$ is used and it is 
\bea
E = e^{3A(r)} \, L^7 \, a(r) \, \sin^6 \theta \, \sin \theta_2 \, \sin^2 \theta_3
\, \sin^3 \theta_4 \, \sin^4 \theta_5 \, \sin^5 \theta_6 \, \left[
a^2(r) \, \cos^2 \theta
+ b^2(r) \, 
\sin^2 \theta \right]^{\frac{2}{3}},
\nonu
\eea
by calculating the determinant of 11-dimensional metric (\ref{11met})
with (\ref{met1}).

Therefore, the ${\cal N}=1$ $G_2$-invariant 
solutions (\ref{fst1}), with the Appendix $A$ where 
the Ricci tensor is presented and the Appendix $B$ where the
coefficient functions are listed,  satisfy the field equations 
(\ref{fieldequations}) as long as the deformation parameters $(a,b)$
of the 7-ellipsoid (or two supergravity fields in 4-dimensions) 
and amplitude $A$ develop in the $AdS_4$ radial
direction along the $G_2$-invariant flow (\ref{flow}).
The corresponding gauge dual was constructed in the context of 
${\cal N}=1$ $U(2) \times U(2)$ superconformal 
Chern-Simons matter theory by adding one mass term for the adjoint
${\cal N}=1$ superfield and the matter multiplet consists of
seven flavors transforming in the adjoint together with flavor
symmetry ($\bf 7$ of $G_2$) \cite{Ahn0806n1}.   

\subsection{Spherical parametrization II}

Let us describe the other spherical parametrization describing the unit
round five-sphere ${\bf S}^5$ whose base is characterized by ${\bf
  CP}^2$ space
\bea
u^1 + i \, u^2 & = & \sin \theta_4 \, \cos (\frac{\theta_1}{2})  \, e^{\frac{i}{2}
(\theta_2 + \theta_3)}\, e^{i \theta_5},
\nonu \\
u^3 + i \, u^4 & = & \sin \theta_4 \, \sin (\frac{\theta_1}{2})  \, e^{\frac{i}{2}
(-\theta_2 + \theta_3)} \, e^{i \theta_5},
\nonu \\
u^5 + i \, u^6 & = & \cos \theta_4
\,  e^{i \theta_5}.
\label{othersixu}
\eea
The isometry of ${\bf S}^5$ is $SU(3) \times U(1)$ where 
$SU(3)$ acts on three complex coordinates $z^i \equiv u^{2i-1} + i \,
u^{2i}$($i=1, 2, 3$) and $U(1)$ acts on each $z^i$ as the phase rotations. 
The vector $u$ spans the ${\bf S}^5$ given by Hopf fibration 
on ${\bf CP}^2$ space. This can be described by writing $(du)^2$
as $(d u)^2 = ds_{FS(2)}^2 + (u, J d u)^2$ where $ds_{FS(2)}^2$ denotes 
the Fubini-Study metric on ${\bf CP}^2$ and $(u, J d u)$ is the Hopf
fiber on it. The $J$ is the standard Kahler form.
One can introduce the two orthogonal unit vectors (\ref{UV}) and the
unit normal vector (\ref{normal}) for given above parametrization (\ref{othersixu}).
Furthermore, the set of the six basis vectors is given by (\ref{ei}) 
and they as well as the normal vector can be described in terms of
imaginary octonions. 
More explicitly, let us present some of them here
\bea
\hat{\bf e}_1^1  & = & \frac{1}{2} \, \cos (\frac{\theta_2 +\theta_3}{2}
+\theta_5) \, \sin \frac{\theta_1}{2} \, \sin \theta_4 \, \sin \theta_6,
\nonu \\
\hat{\bf e}_1^2  & = & \frac{1}{2} \, \sin \frac{\th_1}{2} \sin \th_4
\, \sin
(\frac{\th_2+\th_3}{2} +\th_5) \, \sin \th_6, 
\qquad \cdots \qquad,
\nonu 
\eea
and the other nonvanishing components are given in the Appendix $C$.

The three $G_2$-invariant tensors $\hat{F}_{ij},
\hat{T}_{ijk}$ and $\hat{S}_{ijk}$ are given 
by (\ref{fts}) for the choice of (\ref{othersixu}).
For example, 
\bea
\hat{F}_{12}  & = & \frac{1}{32} \sin^2 \theta_4 \sin^2 \theta_6
\left[ -4 \cos 2\alpha_2 \cos \theta_6 + 4 \cos 2\alpha_1 \cos
  2(\alpha_3 +\theta_5) \cos \theta_6 \right. \nonu \\
& + & ( 8 \cos \alpha_1 \cos
  (\alpha_2 - \alpha_3-\theta_5) \sin \theta_4 -\cos \theta_4  ( \cos 
(2\alpha_1-2\alpha_3-3\theta_5)\nonu \\
& - &  \cos (2\alpha_1-2\alpha_3 -\theta_5)-
\cos (2\alpha_1 + 2\alpha_3+\theta_5) + \cos (2\alpha_1 +
2\alpha_3+3 \theta_5) \nonu \\
& + &  \left. 8 \cos \alpha_1 \cos \theta_5 \sin \alpha_1 +
8 \cos \alpha_2 \sin \alpha_2 \sin \theta_5)) \sin \theta_6
\right],
 \nonu \\
\hat{F}_{13}  & = & \frac{1}{32} \sin^2 \theta_4 \sin^2 \theta_6 
\left[ -2( \cos 2(\alpha_1 -\alpha_2) + \cos 2(\alpha_1 + 
\alpha_2) -2 \cos 2(\alpha_3 + \theta_5)) \cos \theta_6 
\right. \nonu \\
& + & ( 8 \cos \alpha_1 \cos (\alpha_2-\alpha_3-\theta_5) 
\sin \theta_4 + \cos \theta_4 ( \cos (2\alpha_1-2\alpha_2-
\theta_5) \nonu \\
& - & \cos (2\alpha_1 +2\alpha_2-\theta_5)
-\cos (2 \alpha_1 -2\alpha_2 +\theta_5) + \cos 
(2\alpha_1 +2\alpha_2 +\theta_5) 
\nonu \\
&+& \left. 8  \cos \alpha_1 \cos \theta_5 \sin \alpha_1 + 4 \sin \theta_5
\sin 2(\alpha_3 + \theta_5))) \sin \theta_6 \right],
\qquad \cdots \qquad.
\nonu 
\eea
These tensors are given in the Appendix $D$ in terms of internal
coordinates $\theta_i$($i=1, \cdots, 6$) explicitly.
Note that the antisymmetric tensor on ${\bf S}^6$, the dual of $T_{ijk}$, 
has an extra minus sign as follows:
\bea
\hat{S}_{ijk} = -\hat{F}_i^{\, l} \, \hat{T}_{ljk}.
\label{cp2condition}
\eea
All the identities between three tensors $\hat{F}_{ij}, \hat{T}_{ijk}$
and 
$\hat{S}_{ijk}$ are given in the Appendix $D$.
 
By computing the dot product between $\hat{\bf e}_i$ and $\hat{\bf e}_j$
in ${\bf I}^7$, one obtains the 6-dimensional metric on ${\bf S}^6$.
That is,
\bea
g_{ij} =
\left(
\begin{array}{cccccc}
\frac{1}{4} s_4^2 \, s_6^2 & 0 &0 &0 &0 &0   \\ 
0 & \frac{1}{4} s_4^2 \, s_6^2  & \frac{1}{4} c_1 \, s_4^2 \, s_6^2 &0  
&  \frac{1}{2} c_1 \, s_4^2 \, s_6^2 & 0   \\
0 & \frac{1}{4} c_1 \, s_4^2 \, s_6^2 
&  \frac{1}{4} s_4^2 \, s_6^2  &0 &  \frac{1}{2}  s_4^2 \, s_6^2 &
0  \\
0 & 0 &0 & s_6^2 & 0 & 0  \\
0 & \frac{1}{2} c_1 \, s_4^2 \, s_6^2 & \frac{1}{2}  s_4^2 \, s_6^2 & 0 & s_6^2  & 0 \\
0 & 0 &0 & 0 & 0 &  1
\end{array} \right), c_i \equiv \cos \th_i, \qquad s_i \equiv \sin \th_i.
\label{Met6d}
\eea
One sees this 6-dimensional metric from the projection of
7-dimensional metric $g_{mn}^7$ \cite{AW0907} by putting the extra coordinate
$\theta$ to $\frac{\pi}{2}$.
The rectangular coordinates parametrizing the ${\bf S}^7$ inside ${\bf
R}^8$ with ${\bf S}^6 \simeq G_2/SU(3)$ base are given by introducing
the seventh coordinate $\theta$ \cite{AI02}
\bea
X^i = u^i \, \sin \theta_6 \, \sin \theta, \qquad 
X^7= \cos \theta, \qquad X^8 =\cos \theta_6 \, \sin \theta,
\label{rect}
\eea 
with $\sum_{A=1}^{8} (X^A)^2 =1$. 
As done in previous spherical parametrization, via the nonlinear
metric ansatz \cite{dWNW}, the warp factor and the 7-dimensional inverse 
metric are completely determined.
Then the  7-dimensional warped ellipsoidal 
metric in this case can be summarized as \cite{dWNW,AI02}
\bea
d s_7^2 =   \sqrt{\Delta \, a} \, L^2 \left[ \frac{\xi^2}{a^2} \, d
  \theta^2 +
\sin^2 \theta  \, d \Omega_6^2 \right], \qquad
d \Omega_6^2 =  d \theta_6^2 + \sin^2 \theta_6 \left[ d
    s_{FS(2)}^2 + (u, J d u)^2 \right]. 
\label{met}
\eea
The quadratic form and warp factor are given by (\ref{xi2}) and
(\ref{warp}) 
respectively. 
The fact that the vector $(u^1, u^2, u^3, u^4, u^5, u^6)$ in (\ref{othersixu}) spans
the ${\bf S}^5$ with ${\bf CP}^2$ base, compared to the one in
(\ref{met1}) that looks similar to (\ref{met}) but the 5-dimensional
metric inside behaves differently, 
can be  
understood by writing $(d u)^2$ in terms of the Fubini-Study metric
\bea
 d s_{FS(2)}^2 = d \theta_4^2 + \frac{1}{4} \, \sin^2 \theta_4
 \left(\sigma_1^2 + \sigma_2^2 + \cos^2 \theta_4 \, \sigma_3^2 \right),
\label{fs2} 
\eea
and its Hopf fiber
\bea
(u, J d u) = d \theta_5 + \frac{1}{2} \, \sin^2 \theta_4 \, \sigma_3. 
\label{ujdu}
\eea
The standard Kahler form $J$ is given by 
$J_{12} =J_{34}=J_{56}=J_{78}=1$.
One can easily check that the 6-dimensional metric (\ref{met}) is nothing but 
the metric in (\ref{Met6d}) and the one-forms are given by 
$\si_1 = \cos \th_3 d \th_1 + \sin \th_1 \sin \th_3 d \th_2,
\si_2 = \sin \th_3 d \th_1 - \sin \th_1 \cos \th_3 d \th_2,$ and 
$\si_3 = d \th_3 + \cos \th_1  d \th_2$, as usual.
Recall that the $G_2$ symmetry is the isometry group of the metric.
The Killing vector associated to the $G_2$ symmetry is constructed
explicitly in \cite{AW0907} and it is shown that 
the 7-dimensional metric of (\ref{met}) has vanishing Lie derivative \cite{DNP}
with respect to the Killing vector fields. 

The 11-dimensional metric is realized as (\ref{11met}) together with
(\ref{met}). The nontrivial task is to find the exact solution for
11-dimensional Einstein-Maxwell equations (\ref{fieldequations}).
Also we assume 
the nontrivial $AdS_4$ radial coordinate dependence of vacuum expectation
values and they are subject to the 4-dimensional RG flow equations
(\ref{flow}).
The Ricci tensor can be computed from the 11-dimensional metric with 
(\ref{11met}) together with (\ref{met})  directly and 
they have the same form as the ones in previous spherical
parametrization.
So they have common feature in the Ricci tensor presented in the
Appendix $A$. This indicates that the Ricci tensor is indeed independent of
the particular 5-dimensional space.
One can express the Ricci tensor in the frame basis rather than in the
coordinate basis. The former is exactly the same as the latter except
the off-diagonal $(4,5)$- and $(5,4)$-components.

It is natural to solve the 11-dimensional Einstein-Maxwell equations
by assuming that the 4-forms have the same tensorial structure as
the ones in (\ref{fst1}) and these $G_2$-invariant tensors are 
defined in the present 6-dimensional metric (\ref{met}).
The undetermined coefficient functions depend on the coordinates 
$(r, \theta)$. One should find out these coefficient functions
explicitly by requiring that 4-forms should satisfy (\ref{fieldequations}). 
The field strengths are given by
\begin{eqnarray}
F_{\mu\nu\rho 4} & = & e^{3A(r)}
\,W_r (r,\theta)\,\epsilon_{\mu\nu\rho},\qquad
F_{\mu\nu\rho 5} = e^{3A(r)}
\,W_{\theta}(r,\theta)\,\epsilon_{\mu\nu\rho},
\nonu \\
F_{mnpq} &=& 2\,h_2 (r,\theta)\,\eps6_{mnpqrs} \, \hat{F}^{rs}, \qquad
F_{5mnp} = \tilde{h}_1 (r,\theta)\, \hat{T}_{mnp}
+\tilde{h}_2 (r,\theta)\, \hat{S}_{mnp}, \nonumber\\
F_{4mnp} &=& \tilde{h}_3 (r,\theta)\, \hat{T}_{mnp}
+\tilde{h}_4 (r,\theta)\, \hat{S}_{mnp}, \qquad
F_{45mn} = \tilde{h}_5 (r,\theta)\, \hat{F}_{mn}, 
\label{fst2}
\end{eqnarray}
where we denote the three hatted tensors here in order not to confuse
with the unhatted ones that appear in (\ref{fst1}).  
Applying these field strengths (\ref{fst2}) to the 11-dimensional Einstein-Maxwell
equations, one can determine the unknown coefficient functions, which
depend on $r$ and $\theta$, completely. 
If we compute the right hand side of Einstein equation, then 
the linear combination of quadratic coefficient functions appears 
after using the identities between $G_2$-invariant tensors or
calculating the functional expressions component by component explicitly. 
For latter case, we do not need to use the identities because 
we know  three $G_2$-invariant tensors according to the Appendix $D$ 
\footnote{One can easily check that 
the above rank 2 and rank 3 tensors written in terms of the angle variables
can be expressed similarly as in footnote 1.
That is,  $\hat{F}^{(2)} = \frac{
1 }{(\sum_{A=1}^7 (X^A)^2)^{\frac{3}{2}}} \, \eta_{ABC} \, X^A d X^B \wedge d X^c$,
$\hat{S}^{(3)} = -\frac{
1 }{(\sum_{A=1}^7 (X^A)^2)^2} \, \eta_{ABCD} \, X^A d X^B \wedge d X^c \wedge d X^D$, and
$\hat{T}^{(3)} = d \hat{F}^{(2)}$. Note the extra minus sign in the
3-form $\hat{S}^{(3)}$. As in footnote 1, we interchange
the $X^7$ and the $X^8$ each other. It is not so obvious to express
the angular variables in terms of the rectangular coordinates,
contrary to the footnote 1 and from the begining we assume the above
forms.
After inserting the rectangular coordinates written in terms of
angular variables (\ref{othersixu}) and (\ref{rect})  
into them, we have checked
these agree with those in the Appendix $D$. Therefore, the
$G_2$-invariance is more obvious in this basis rather than in the
angle basis. Note that the $G_2$-invariant tensors $\hat{F}_{mn},
\hat{T}_{mnp}$ and 
$\hat{S}_{mnp}$ in (\ref{fst2}) 
can be obtained from those in previous parametrization
via the coordinate transformations between (\ref{sixu}) and
(\ref{othersixu}).
In particular, we have checked that the component $F_{46}$ is related to  the
component $\hat{F}_{46}$ via coordinate transformations.}.  
Except $(4,5)$- and $(5,4)$-components of Einstein equations, 
$(1,1), (4,4), (5,5)$- and
$(6,6)$-components depend on eight coefficient functions. 
The right hand side of Einstein equations can be summarized by
(3.21) of \cite{AI} exactly. As we explained before, 
the left hand side of Einstein equations, the Ricci tensor, 
is identical to each other. Therefore, one concludes that
the unknown coefficient functions  are
exactly the same as those in \cite{AI} or in the Appendix $B$.
Note that 
the geometric superpotential $\widetilde{W}$ from the field strengths
$(W_r = e^{-3A} \pa_r (e^{3A} \widetilde{W}), 
W_{\theta} = \pa_{\th}  \widetilde{W})$ 
was found by requiring that it should coincide with   
the 4-dimensional superpotential $W$ (\ref{spoten}) up to a multiplicative constant
when the internal coordinate $\theta$ is fixed to some specific value.

The analysis for the checking of Maxwell equations can be done before 
exactly, based on the paragraph containing (\ref{max}). 
We have checked the solutions (\ref{fst2}) 
also satisfy the Maxwell equations explicitly. 
During this computation, 
the elfbein determinant $E=\sqrt{-g_{11}}$ from 11-dimensional metric
(\ref{11met}) and (\ref{met}) is 
given by
\bea
E = \frac{1}{8} \, e^{3A} \, L^7 \, a \, 
\sin^6 \theta \, \sin \theta_1 \, \cos \theta_4 
\, \sin^3 \theta_4  \, \sin^5 \theta_6 \, \left[
a^2 \, \cos^2 \theta
+ b^2 \, 
\sin^2 \theta \right]^{\frac{2}{3}}.
\nonu
\eea

Therefore, we have established that the solutions (\ref{fst2})
together with the Appendices $A, B$ and $D$
consists of an exact solution to 11-dimensional supergravity by
bosonic field equations (\ref{fieldequations}), provided that the
deformation parameters $(a, b)$ of the 7-dimensional internal
space and the domain wall amplitude $A$ develop in the $AdS_4$
radial direction along the $G_2$-invariant RG flow (\ref{flow}).
Although the 11-dimensional metric and 4-forms are different 
from those in previous parametrization, the Ricci tensor (or the
quadratic structure of 4-forms appearing in the right hand side of
Einstein equations) is exactly the same each other.  

\section{Towards an ${\cal N}=1$ $SU(3)$-invariant 
supersymmetric flow in an 11-dimensional theory}

We will use the results of 
11-dimensional solutions with ${\bf CP}^2$ in previous subsection.
In the context of the round seven-sphere ${\bf S}^7$ as Hopf fibration
on ${\bf CP}^3$ let us 
replace the coordinate $\theta_5$ in
(\ref{othersixu}) with $\phi+\psi$ where $\psi$ is the coordinate of
the Hopf fiber on ${\bf S}^7$ as follows:
\bea
u^1 + i \, u^2 & = & \sin \theta_4 \, \cos (\frac{\theta_1}{2})  
\, e^{\frac{i}{2}
(\theta_2 + \theta_3)}\, e^{i (\phi+\psi)},
\nonu \\
u^3 + i \, u^4 & = & \sin \theta_4 \, \sin (\frac{\theta_1}{2})  
\, e^{\frac{i}{2}
(-\theta_2 + \theta_3)} \, e^{i (\phi+ \psi)},
\nonu \\
u^5 + i \, u^6 & = & \cos \theta_4
\,  e^{i (\phi + \psi)}.
\label{us} 
\eea
As before, the vector $(u^1, u^2, u^3, u^4, u^5, u^6)$ in (\ref{us}) spans
the ${\bf S}^5$ with ${\bf CP}^2$ base and the Fubini-Study metric on
${\bf CP}^2$ is
given by (\ref{fs2}).
The correct 7-dimensional warped metric, corresponding to (5.15) of 
\cite{AI02}  is given by 
\bea
d s_7^2 = \sqrt{\Delta \, a} \, L^2 \left[ (d U)^2 + \frac{b^2}{a^2}
  \, (d V_1)^2  + (d V_2)^2 \right],
\label{metmet}
\eea
where the ${\bf R}^8$ coordinates $X^A$ are decomposed into
\bea
{\bf U} & = & (u^1, u^2, u^3, u^4, u^5, u^6, 0, 0) \, \cos \mu, \nonu \\
{\bf V}_1  & = & (0, 0, 0, 0, 0, 0, 1, 0) \, \cos \psi \, \sin \mu, \nonu \\
{\bf V}_2  & = & (0, 0, 0, 0, 0, 0, 0, 1) \, \sin \psi \, \sin \mu.
\label{UV1V2}
\eea
The deformation parameter $\gamma \equiv \frac{c d}{a b} -1$ 
of \cite{AI02} vanishes for the $G_2$-invariant sector in which the
constraints are characterized by $c=a$
and $d=b$.
For $G_2$-invariant sector, the ellipsoidal deformation arises along
the $V_1$ direction. Then one can choose the two additional orthonormal frames as 
$(u, J d u)$ (\ref{ujdu}) and $dV_1$ respectively. Then the remaining orthonormal frame 
is determined from (\ref{metmet}) by completing squares.

The precise relations between the parameters of previous section and
those in this section by comparing (\ref{rect}) with
(\ref{UV1V2}) (other eight coordinates are the same) are given by
\bea
\cos \theta & = & \cos \psi \, \sin \mu, \nonu \\
\theta_5 & = & \phi + \psi,
\nonu \\
\cos \theta_6 & = & \frac{\sin \mu \, \sin \psi}{\sqrt{1-\cos^2 \psi \,
    \sin^2 \mu}}.
\label{rel}
\eea
From these (\ref{rel}), one obtains the partial differentiations of
old variables $(\theta, \theta_5, \theta_6)$ with
respect to the new variables $(\mu, \phi, \psi)$ as follows:
\bea
\frac{\pa \theta}{\pa \mu}  & = & -\frac{\cos \mu \, \cos \psi }{\sqrt{1-\cos^2 \psi \,
    \sin^2 \mu}}, \qquad
\frac{\pa \theta}{\pa \psi}   =  \frac{\sin \mu \, \sin \psi }{\sqrt{1-\cos^2 \psi \,
    \sin^2 \mu}},
\nonu \\
\frac{\pa \theta_5}{\pa \phi}  & = & 1, \qquad
\frac{\pa \theta_5}{\pa \psi}   =  1,
\nonu \\
\frac{\pa \theta_6}{\pa \mu}  & = & -\frac{\sin \psi }{1-\cos^2 \psi \,
    \sin^2 \mu}, \qquad
\frac{\pa \theta_6}{\pa \psi}   =  -\frac{\sin \mu \, \cos \mu \, 
\cos \psi }{1-\cos^2 \psi \,
    \sin^2 \mu}.
\label{chain} 
\eea

The frames for the 11-dimensional metric is summarized by
\bea
e^1  & = & \Delta^{-\frac{1}{2}} \, e^A \, d x^1,
\qquad
e^2   =  \Delta^{-\frac{1}{2}} \, e^A \, d x^2,
\qquad
e^3   =  \Delta^{-\frac{1}{2}} \, e^A \, d x^3,
\qquad
e^4    =   \Delta^{-\frac{1}{2}}  \, d r,
\nonu \\
e^5 & = & {L} \,
\Delta^{\frac{1}{4}} \,   a^{\frac{1}{4}}  \,
\frac{\xi}{a} \, \sqrt{\frac{1}{1-\cos^2 \psi \,
    \sin^2 \mu}} \,
\left(-
\cos \mu \, \cos \psi \, d \mu + \sin \mu \, \sin \psi \, d \psi \right),
\nonu \\
 e^6  &  = &  {L} \,
\Delta^{\frac{1}{4}} \, a^{\frac{1}{4}} \,  \cos \mu \, 
\frac{1}{2} \, \sin \theta_4 \, \sigma_1,
\nonu \\
 e^7 & = & {L} \,
\Delta^{\frac{1}{4}} \,  a^{\frac{1}{4}} \, \cos \mu \, 
\frac{1}{2} \, \sin \theta_4 \, \sigma_2,
\nonu \\
 e^8  & = &  {L} \,
\Delta^{\frac{1}{4}} \, a^{\frac{1}{4}} \, \cos \mu \, 
\frac{1}{2} \, \sin \theta_4 \,\cos \theta_4 \, \sigma_3,
\nonu \\
 e^9 & = & {L} \,
\Delta^{\frac{1}{4}} \,  a^{\frac{1}{4}} \, \cos \mu \, 
d \theta_4,
\nonu \\
 e^{10} & = & {L} \,
\Delta^{\frac{1}{4}} \,  a^{\frac{1}{4}} \,
\cos \mu \, 
\left( d \phi + d \psi
  + \frac{1}{2} \, \sin^2 \theta_4 \, \sigma_3
 \right),
\nonu \\
e^{11}   & = &  {L} \,
\Delta^{\frac{1}{4}} \,  a^{\frac{1}{4}} \,  
 \sqrt{\frac{1}{1-\cos^2 \psi \,
    \sin^2 \mu}} \,
\left(-
\sin \psi \, d \mu - \sin \mu \, \cos \mu \, \cos \psi \, d \psi \right).
\label{11frames}
\eea
The $\widetilde{e}^5, \widetilde{e}^6$ and $\widetilde{e}^7$ with some
sign difference appearing
in (5.24)  of \cite{AI02}
correspond to $e^{10}, e^{11}$ and $e^{5}$ respectively.
The frames $e^6, e^7, e^8$ and $e^9$ correspond to the Fubini-Study
metric on ${\bf CP}^2$ whose symmetry group is $SU(3)$. 
Note that  $\psi$ which is the coordinate of
the Hopf fiber on ${\bf S}^7$ appears in the frame $e^{10}$ and
this gives the off-diagonal metric components between the coordinates $(\psi,
\th_2,\th_3, \phi)$. Also from the structure of $e^5$ and $e^{11}$, 
the off-diagonal components between $(\mu,\psi)$ occur.
The quadratic form (\ref{xi2}) reads  as $ \xi^2 =
a^2 \cos^2 \psi \, \sin^2 \mu 
+ b^2 \, (1-\cos^2 \psi \sin^2 \mu )$ from (\ref{rel}).
The warp factor is the same as before: (\ref{warp}). The one-forms 
$\si_i$ are given as before \footnote{When we take the seven
  rectangular coordinates except $X^7$ among (\ref{UV1V2}), the
  similar computation as in \cite{AW0907} leads to the fact that
the 7-dimensional metric has vanishing Lie
  derivative with respect to the Killing vector fields
where the generator for $G_2$ is the same as the one in
  the Appendix $A$ of \cite{AW0907}. This implies that the Killing
  vector is associated to the $G_2$ symmetry. }. 

Then how does one find the solution for 11-dimensional
Einstein-Maxwell equations for given 11-dimensional background
(\ref{11frames})? 
The 6-dimensional space is no longer unit round six-sphere.
Since the metric (\ref{met}) is related to the metric (\ref{11frames}) by
change of variables, one can use the solutions (\ref{fst2}) and find
out the  solutions. 
First of all, the Ricci tensor can be obtained from (\ref{11frames})
directly
or can be determined from the one in previous section by using the
transformation on the coordinates between the two coordinate
systems (\ref{chain}) via tensorial property.
In other words, the Ricci tensor is given by explicitly
\bea
\widetilde{R}_{M}^{\,\,N} =  \left(\frac{\pa z^P}{ \pa \widetilde{z}^{ M}} \right)
\left(\frac{\pa \widetilde{z}^{ N}}{\pa z^Q}
\right) R_{P}^{\,\,Q},
\label{rtilde}
\eea
where 
the 11-dimensional coordinates are given by
\bea
z^M   =  (x^1, x^2, x^3, r; \theta, \theta_1, \theta_2, 
\theta_3, \theta_4, \theta_5, \theta_6), \qquad
\widetilde{z}^M   =  (x^1, x^2, x^3, r; \mu, \theta_1, \theta_2, 
\theta_3, \theta_4, \phi, \psi).
\nonu 
\eea 
Eight of them are common and three of them are distinct.
The Ricci tensor $R_{P}^{\,\,Q}$ for $G_2$-invariant flow
is presented in the Appendix $A$.
Some of them are given by
\bea
\widetilde{R}_{1}^{\,\,1} & = & R_{1}^{\,\,1}, \qquad
\widetilde{R}_{2}^{\,\,2}  =  R_{2}^{\,\,2}\ (= \widetilde{R}_{1}^{\,\,1}), \qquad
\widetilde{R}_{3}^{\,\,3} =  R_{3}^{\,\,3} (= \widetilde{R}_{1}^{\,\,1}),  
\qquad \cdots \qquad . 
\nonu 
\eea
There exist off-diagonal components
$
(4,10), (4,11), (5,10), (5,11), 
(11,4), (11,5), (11,10)$,
as well as $(4,5)$- and $(5,4)$-components.
Their full expressions  are given in the Appendix $E$.

For the field strengths, one has, by multiplying the transformation
matix,
\bea
\widetilde{F}_{MNPQ} = \left(\frac{\pa z^R}{ \pa \widetilde{z}^{M}} \right)
\left(\frac{\pa z^S}{\pa \widetilde{z}^{ N}}
\right) \left( \frac{\pa
  z^T}{\pa \widetilde{z}^{ P}}\right) \left(\frac{\pa z^U}{\pa
  \widetilde{z}^{ Q}}
\right) F_{RSTU}.
\label{ftilde}
\eea
Then some of the components, using the relations (\ref{chain}), are 
given by
\bea
\widetilde{F}_{1234} & = & F_{1234}, \qquad
\widetilde{F}_{1235}   =  -\left[\frac{\cos \mu \, \cos \psi}
{\sqrt{1-\cos^2 \psi \, \sin^2
    \mu}} \right] F_{1235}, 
\qquad \cdots \qquad .
\nonu 
\eea
These transformed 4-forms  
are given in terms of those in $G_2$-invariant flow in
the Appendix $F$ and moreover, the transformed 4-forms with upper indices are
described.
The 4-form $\widetilde{F}_{123\,11}$ is a new object.
At the IR critical point where $a= \sqrt{\frac{6\sqrt{3}}{5}}$ and 
$b=\sqrt{\frac{2\sqrt{3}}{5}}$, due to the fact that 
the coefficient functions $W_{\theta}, \widetilde{h}_3,
\widetilde{h}_4$ 
and $\widetilde{h}_5$ in the Appendix $B$ vanish, 
the following 4-forms also vanish at this critical point:
\bea
\widetilde{F}_{1235} =0 = \widetilde{F}_{123\, 11} =
\widetilde{F}_{45mn} =\widetilde{F}_{4mnp}.
\nonu
\eea
Note that for the $G_2$-invariant flow, the 4-forms $F_{1235}, F_{4mnp}$ and
$F_{45mn}$ vanish
where $m, n, p=6, \cdots, 11$ at the IR critical point.
Once we suppose that 4-dimensional metric has the domain wall factor
$e^{3A(r)}$ which breaks the 4-dimensional conformal invariance, 
the mixed 4-forms occur along the whole RG flow.
Of course, at the UV critical point where $a=1=b$, the only nonzero
4-form field is $\widetilde{F}_{1234}$.
Some of the transformed 4-forms with upper indices are given by
\bea
\widetilde{F}^{1234} & = & F^{1234}, \qquad
\widetilde{F}^{1235}   =  -\left[\frac{\cos \mu \, \cos \psi}
{\sqrt{1-\cos^2 \psi \, \sin^2
    \mu}} \right] F^{1235}, \qquad \cdots \qquad.
\nonu 
\eea
These can be obtained from (\ref{ftilde}) by using the 11-dimensional
metric (\ref{11frames}) 
or by multiplying the transformation matrices (\ref{chain})
into the 4-forms with upper indices of $G_2$-invariant flow.
The remaining 4-forms are given by the Appendix $F$ explicitly
\footnote{One can also check that 
the field strengths written in terms of the angle variables
can be expressed similarly as in footnotes 1 and 2: 
$\widetilde{F}^{(4)} = d A^{(3)} + h_2 \, d \hat{S}^{(3)}+
d r \wedge (\tilde{h}_3 \, d \hat{F}^{(2)} + \tilde{h}_4 \, \hat{S}^{(3)})
+\frac{1}{\sqrt{1-(X^8)^2}}
d X^8 \wedge (\tilde{h}_1 \, d \hat{F}^{(2)} + \tilde{h}_2 \, \hat{S}^{(3)}) -
\frac{\tilde{h}_5}{\sqrt{1-(X^8)^2}} d r \wedge \hat{F}^{(2)} \wedge d
X^8$ where the 2-form $\hat{F}^{(2)}$ and the 3-forms $A^{(3)}$ and
$\hat{S}^{(3)}$ are given in the footnote 2.
As in footnote 2, it is not so obvious to express
the angular variables in terms of the rectangular coordinates,
we assume the above forms.
After inserting the rectangular coordinates written in terms of
angular variables (\ref{us}) and (\ref{UV1V2})  
into them, we have checked
this 4-form agrees with those in the Appendix $F$. Therefore, the
4-form has $G_2$-invariance explicitly.}.  

How does one check the 11-dimensional solution for Einstein equation?
One shows this by substituting the solutions (\ref{rtilde}) and
(\ref{ftilde})
into the first equation of (\ref{fieldequations}) with transformed-Ricci
tensor and transformed-4-forms. Or one checks this equality using the solution of
$G_2$-invariant flow in subsection 2.2. 
In previous section we have shown that (\ref{fst2})
satisfy the field equations (\ref{fieldequations}). Let us go back the
transformed-Ricci tensor (\ref{rtilde}) which is written in terms of the
Ricci tensor for $G_2$ invariant flow. 
Without specifying the explicit form for Ricci tensor $R_{P}^{\,\,Q}$,
one writes down transformed-Ricci tensor, as in the Appendix $E$. Now one can
use the property of $G_2$-invariant flow: the first equation of 
(\ref{fieldequations}). That is, one replaces the Ricci tensor in
terms of the quadratic 4-forms. Then one sees the transformed-Ricci tensor 
can be written in terms of the quadratic 4-forms for $G_2$-invariant flow. 
Let us return to the right hand side of Einstein equations.      
Using (\ref{ftilde}) and (\ref{chain}), one can express this 
in terms of quadratic 4-forms for $G_2$-invariant flow.

So far, we did not insert the explicit form for the 4-forms $F_{MNPQ}$.
One can make the difference between the left hand side and the right
hand side of Einstein equations and see whether this is zero or not. 
At first sight, some of the
components written in terms of quadratic 4-forms in $G_2$-invariant flow 
are not exactly vanishing. They contain the terms $F_{4npq} \, F^{mnpq}$, 
$F_{5npq} \, F^{mnpq}$, $F_{mnpq} \, F^{snpq}$ where $m,n,p,q = 6,
\cdots, 11$ and $s=4, 5, 6, \cdots, 11$.
After plugging the explicit solution of 
$G_2$-invariant flow, then all of these are equal to zero identically.  
One can also check this from the useful identities between three-invariant
tensors $\hat{F}_{ij}, \hat{T}_{ijk}$ and $\hat{S}_{ijk}$  
presented in the Appendix $D$, as done in \cite{AI}.
Recall that these quadratic 4-forms above 
correspond to the off-diagonal terms of Einstein
equation for $G_2$-invariant flow which vanish identically except
$(4,5)$- and $(5,4)$-components. Of course, 
the above extra piece does not possess
these nonzero off-diagonal terms, $(4,5)$- and $(5,4)$-components. 
Therefore, we have shown the solutions (\ref{rtilde}) and (\ref{ftilde})
indeed satisfy the 11-dimensional Einstein equations. 

Let us move on the Maxwell equations.
Let us introduce the notation 
$
\frac{1}{2} \, E \,  
\widetilde{\nabla}_M  \, \widetilde{F}^{MNPQ} \equiv
(\widetilde{N}\widetilde{P}\widetilde{Q})$ for simplicity,
and present all the nonzero components of left hand side of Maxwell
equations in terms of the 4-forms in $G_2$-invariant flow, using the
property of 11-dimensional solution we have found before.
For example, the $(\widetilde{1}\widetilde{2}\widetilde{3})$-component 
of the left hand side of 
Maxwell equations reads 
\bea
(\widetilde{1}\widetilde{2}\widetilde{3}) & = & 
\sum_{m,n,p,q,r,s=6}^{11} \eps6_{mnpqrs} \,
\left(\frac{1}{3!\, 3!} F_{4mnp}\, F_{5qrs} -\frac{1}{2!\, 4!} 
F_{45mn} \, F_{pqrs} \right).
\nonu 
\eea
The other nonzero components are given in the Appendix $G$ explicitly.
The nonzero components of the Maxwell equation, the second equation of
(\ref{fieldequations}), are characterized by the following indices 
\bea
(\widetilde{1}\widetilde{2}\widetilde{3}), 
\qquad (\widetilde{4}\widetilde{5}\widetilde{m}), 
\qquad  (\widetilde{4}\widetilde{n}\widetilde{p}), 
\qquad (\widetilde{5}\widetilde{n}\widetilde{p}), 
\qquad (\widetilde{m}\widetilde{n}\widetilde{p}), 
\qquad  \widetilde{m}, \widetilde{n} ,\widetilde{p} =6,
\cdots, 11,
\nonu
\eea
with the number of components $1$, $5$ by choosing one out of five(the
(45\,11)-component is equal to zero), 
$15$ by choosing two out of six, 
$15$ by choosing two out of six and $20$ by choosing three out of six respectively. 
Other remaining components of the Maxwell equation become identically zero. 
Therefore, there exist 56-nonzero-components of the Maxwell equation. 
Compared with the ones (\ref{max}) in $G_2$-invariant flow, there are nonzero
terms from $(\widetilde{4}\widetilde{5}\widetilde{m})$-components 
where $\widetilde{m}=6, \cdots, 10$.
The right hand side of Maxwell equations for the 
$(\widetilde{1}\widetilde{2}\widetilde{3})$-component above
consists of 35-terms coming from the quadratic 4-forms.
For the other components, the right hand side contains
a single term, two terms or three terms  in quadratic 4-forms.

Now it is ready to check the Maxwell equations for the solutions 
(\ref{ftilde}). For given 4-forms $\widetilde{F}_{MNPQ}$, 
one can construct the corresponding 4-forms with upper indices
$\widetilde{F}^{RSTU}$ by using the 11-dimensional metric (\ref{11frames}) or
by multiplying the transformation matrix with the 4-forms $F^{MNPQ}$
for $G_2$-invariant flow, as done in (\ref{ftilde}).
Then one obtains all these transformed 4-forms with upper indices, as in the
Appendix $E$. As done in previous paragraph, 
one constructs the left hand side of Maxwell equations in terms of
quadratic 4-forms for $G_2$-invariant flow by using the 11-dimensional
field equations. According to the transformation rules, one can
express the covariant derivative $\widetilde{\nabla}_M$ 
and 4-forms $\widetilde{F}^{MNPQ}$ in terms of $\nabla_N$ and $F^{RSTU}$
for $G_2$-invariant flow together with 
$(\mu,\psi)$-dependent functions.
Then using the Maxwell equations for $G_2$-invariant flow, 
one can replace $\nabla_M \, F^{MNPQ}$ with quadratic 4-forms and
arrives at the Appendix $G$. Similarly, 
let us return to the right hand side of Maxwell equations.      
Using (\ref{ftilde}) and 11-dimensional metric, one can express this 
in terms of quadratic 4-forms in $G_2$-invariant flow.
We also transform the 11-dimensional determinant according to the
transformation rules appropriately. 
It turns out that
the difference between the left hand side and the right
hand side of Maxwell equations becomes zero identically. Therefore, we have
shown that the solution (\ref{ftilde}) indeed satisfies the Maxwell
equations correctly.
During this check, 
the elfbein determinant $E=\sqrt{-g_{11}}$ is used and it is 
\bea
E  & = &  -\frac{1}{8}  e^{3A}  L^7  a  
\sin \mu  \cos^5 \mu  \sin \theta_1  \cos \theta_4 
 \sin^3 \theta_4  
\left[
a^2 \cos^2 \psi  \sin^2 \mu 
+ b^2  (1-\cos^2 \psi \sin^2 \mu ) \right]^{\frac{2}{3}},
\nonu
\eea
by computing the determinant of 11-dimensional metric (\ref{11frames}).

Therefore, we have shown that the solutions (\ref{ftilde})
together with the Appendices $A, B, D, E$ and $F$
consists of an exact solution to 11-dimensional supergravity by
bosonic field equations (\ref{fieldequations}), provided that the
deformation parameters $(a ,b)$ of the 7-dimensional internal
space and the domain wall amplitude $A$ develop in the $AdS_4$
radial direction along the $G_2$-invariant RG flow (\ref{flow})
connecting from ${\cal N}=8$ $SO(8)$ UV fixed point to ${\cal N}=1$
$G_2$ IR fixed point.
Compared with the previous  solutions for $G_2$-invariant flow, 
they share the common
${\bf CP}^2$
space inside 7-dimensional internal space but three remaining
coordinates
are different from each other. 


\section{Conclusions and outlook }

We have found an  exact solution of ${\cal N}=1$ $G_2$-invariant flow (connecting 
from ${\cal N}=8$ $SO(8)$ UV invariant fixed point to ${\cal N}=1$ 
$G_2$ IR invariant fixed point) 
to the 11-dimensional Einstein-Maxwell
equations with (\ref{11met}) and (\ref{met}). Based on this solution, 
we also have discovered an  exact solution of ${\cal N}=1$
$G_2$-invariant 
flow connecting above two fixed points 
to the 11-dimensional Einstein-Maxwell
equations with (\ref{11frames}). 
Now it is ready to look at the 
the 11-dimensional lift of holographic ${\cal N}=1$
supersymmetric $SU(3)$-invariant RG flow \cite{BHPW} 
connecting from ${\cal N}=1$ $G_2$ critical point to 
${\cal N}=2$ $SU(3) \times U(1)_R$ critical
point. At each critical points, there exists a consistent description for
the 11-dimensional metric with ${\bf CP}^3$-basis and there
exist corresponding 4-forms explicitly and it is an open problem to
discover the 11-dimensional solution along the whole ${\cal N}=1$ RG flow.

$\bullet$ What happens when there exist four supergravity fields $(a,
b, c,d)$ \cite{AI02} in which there exists $SU(3)$-invariant sector ? 
So far, we have concentrated on the $G_2$-invariant sector
where there exist two independent supergravity fields. As explained in
the introduction, the 11-dimensional metric is known and it is an open
problem to construct the correct 4-forms. In particular limit, one has
11-dimensional lift \cite{CPW} of  ${\cal N}=2$ $SU(3) \times U(1)_R$-invariant flow and 
for other limit, one obtains the 11-dimensional lift \cite{AI} of
${\cal N}=1$ $G_2$-invariant flow. At least, the 4-forms should respect the behavior
of these two extreme limits. The decoding of the 4-forms written as
the $SU(3)$-singlet vacuum expectation values in
\cite{dN87} will be useful.
One of these flows will describe the ${\cal N}=1$ $SU(3)$-invariant RG flow 
connecting 
from ${\cal N}=8$ $SO(8)$ UV invariant fixed point 
to ${\cal N}=2$ $SU(3) \times U(1)_R$ 
IR invariant fixed point, by looking at the $SU(3) \times
U(1)_R$-invariant 
sector with two supergravity fields.

$\bullet$ What happens when we replace ${\bf CP}^2$ appearing in
(\ref{met}) or
(\ref{11frames}) with ${\bf CP}^1
\times {\bf CP}^1$? In 5-dimensional space, 
this is equivalent to put $T^{1,1}$ space in (\ref{met}). 
According to the branching of $G_2$ into $SU(2)
\times SU(2)$, one expects that the 11-dimensional solution should
preserve $SU(2) \times SU(2)$ symmetry. It would be interesting to
find out the correct 4-forms for given 11-dimensional 
metric for this case: the uplift of ${\cal N}=1$ $SU(2) \times SU(2)$-invariant
flow. In order to do this direction, 
one needs to look at the structure of 4-forms 
found in this paper, in the frame basis.
It is a nontrivial task to find out the
corresponding gauge dual, as seen in \cite{AW0908}. 
Furthermore, the most general 5-dimensional Sasaki-Einstein space can
be considered and the global symmetries become smaller than $SU(2)
\times SU(2)$ symmetry.

$\bullet$ Any octonion description for the present work?
The automorphisms of the Cayley algebra that leaves one of the
imaginary octonion units  form  a subgroup $SU(3)$ of $G_2$ \cite{GW,GG}. 
For the split octonion algebra, the automorphism group $G_2$ acts on
the basis by an 8-dimensional reducible representation. Two of them
are invariant, three split
octonions transform like triplet, and three complex conjugate
split octonions transform like antitriplet,  under the $SU(3)$ 
subgroup of split $G_2$. 

\vspace{.7cm}

\centerline{\bf Acknowledgments}

This work was supported by the 
National Research Foundation of Korea(NRF) grant 
funded by the Korea government(MEST)(No. 2009-0084601).
CA acknowledges warm hospitality of 
School of Physics, at Korea Institute for Advanced Study(KIAS)
where this work was completed.

\appendix

\renewcommand{\thesection}{\large \bf \mbox{Appendix~}\Alph{section}}
\renewcommand{\theequation}{\Alph{section}\mbox{.}\arabic{equation}}

\section{The Ricci tensor for $G_2$-invariant flow }

The Ricci tensor appearing in (\ref{fieldequations}) for
$G_2$-invariant flow  is given by 
\bea
R_{1}^{\,\,1} & = & -\frac{1}{24 L^2 \left[(a^2 + 7 b^2)^2 -112(a 
  b-1)\right]
(a^2  \cos^2 \theta + b^2  \sin^2
  \theta)^{\frac{8}{3}}} \nonu \\
& \times & \left[ 8 a^{14}  \cos^4 \theta - 1120 
  a^{11}  b \cos^4 \theta + 8  a^{12} b^2 \cos^2 \theta
  (15+13  \cos 2\theta) \right.
\nonu \\
&-& 112  a^9 b^3  \cos^2 \theta (83 + 63  \cos 2\theta) -
8  a^5  b^3 (32 \cos^2 \theta(445 + 437\cos 2\theta)
\nonu \\
&+ & 7b^4(617+148\cos2\theta-
177\cos4\theta) ) + 4a^8 b^2 (4\cos^2 \theta (1139 + 1031\cos
2\theta) \nonu \\
&+& 7b^4(171+192\cos2\theta + 29\cos4\theta)) 
+ a^{10} (896 \cos^4 \theta + b^4 (941 + 1172 \cos 2\theta + 239 \cos 4\theta)) \nonu \\
&+& a^6 (-128 \cos^2 \theta (29+ 27\cos 2\theta)-49 b^8(-221-172\cos
2\theta + \cos 4\theta) \nonu \\
&+& 16 b^4(4123+4776\cos 2\theta + 705\cos 4\theta)) - 8 a^7 b(32
\cos^2 \theta(27 + 29 \cos 2\theta) \nonu \\
&+ & 7b^4 (564\cos 2\theta +
73(7+\cos 4\theta))) - 16 a^3 b (768\cos^2 \theta + 49 b^8(83 +\cos
2\theta) \nonu \\
&+& 16b^4 (549 + 697 \cos 2\theta))\sin^2 \theta 
+ 96 b^2 (-384 + 616 b^4 + 245 b^8) \sin^4 \theta
\nonu \\
&-& 96 a b^3 (-1408 + 1232 b^4 + 343 b^8)\sin^4 \theta + 8 a^2 \sin^2
\theta (4608 \cos^2 \theta -784 b^8(-14 + 5\cos 2\theta) \nonu \\
& +& 16 b^4
(37+1209\cos 2\theta)+2401 b^{12} \sin^2 \theta)
+ 8 a^4 b^2 (4988 + 7824 \cos 2\theta \nonu \\
& + & \left. b^4 (7191 + 212 \cos 2\theta
-3287 \cos 4\theta) + 2868 \cos 4\theta + 343 b^8 (9+ 5 \cos
2\theta)\sin^2 \theta) \right] \nonu \\
& = & R_{2}^{\,\, 2} = R_{3}^{\,\, 3}, 
\nonu \\
R_{4}^{\,\,4} & = &  -\frac{1}{24 L^2 \left[(a^2 + 7 b^2)^2 -112(a 
  b-1)\right]
(a^2  \cos^2 \theta + b^2  \sin^2
  \theta)^{\frac{8}{3}}} \nonu \\
& \times & \left[ 8 a^{14}  \cos^4 \theta + 8  a^{12} b^2 \cos^2 \theta
  (15+13  \cos 2\theta)- 8  a^{11} b  \cos^2 \theta (73 + 67  \cos 2\theta) \right.
\nonu \\
&-& 
4  a^5  b^3 (7b^4(2425+1340 \cos2\theta-
237 \cos4\theta) 
 + 16 (2461 + 2268 \cos 2\theta -25 \cos 4\theta)) \nonu \\ 
& - &2 a^9 b^3 (3527 + 4412 \cos 2\theta + 909 \cos 4\theta) 
+ 2a^8 b^2 (9303+ 11692 \cos 2\theta + 2413 \cos 4\theta 
\nonu \\
& + & 14b^4(171+192\cos2\theta + 29\cos4\theta))
+ a^{10} (896 \cos^4 \theta + b^4 (941 + 1172 \cos 2\theta + 239 \cos 4\theta)) \nonu \\
&-&   4 a^7 b(128
\cos^2 \theta (51 + 40 \cos 2\theta) 
+  b^4 (9281+ 10128 \cos 2\theta + 1367 \cos 4\theta)) \nonu \\
&+& a^6 ( 128 \cos^2 \theta (55 + 57 \cos 2\theta)-49 b^8 (-221
-172\cos 2\theta +  \cos 4\theta) \nonu \\
& + & 8 b^4 (14309 + 14988 \cos 2\theta + 1671 \cos 4\theta)) \nonu \\
& - & 8 a^3 b (24576 \cos^2 \theta + 49 b^8(349 +113 \cos
2\theta) 
+ 64 b^4 (942 + 395 \cos 2\theta))\sin^2 \theta 
\nonu \\
&+& 48 b^2 (1536 + 3248 b^4 + 1519 b^8) \sin^4 \theta
- 48 a b^3 (5632 + 7840 b^4 + 1715 b^8)\sin^4 \theta \nonu \\
& +& 8 a^2 \sin^2
\theta (4608 \cos^2 \theta + b^8(29876 - 8708 \cos 2\theta)  +  b^4
(42832- 6768\cos 2\theta)\nonu \\
& + & 2401 b^{12} \sin^2 \theta)
 - 4 a^4 b^2 (8(-2981 - 1884 \cos 2\theta + 945 \cos 4\theta) \nonu \\
& + & \left. b^4 (-36429 + 524 \cos 2\theta
+11209 \cos 4\theta) - 686 b^8 (9+ 5 \cos
2\theta)\sin^2 \theta) \right], 
\nonu \\
%
R_{4}^{\,\,5} & = &  \frac{1}{8 L^3 \sqrt{(a^2 + 7 b^2)^2 -112(a 
  b-1)}
(a^2  \cos^2 \theta + b^2  \sin^2
  \theta)^{\frac{8}{3}}} 
\nonu \\
& \times &  \left[ a^{\frac{5}{2}} (-8a^5 (80 + a^4) \cos^3 \theta
  \sin \theta + 4 b (a^2 ( 48 ( 4+a^4) -a ( 192 +7 a^4) b \right. \nonu \\
& + & 54 a^2 b^2 -24 a^3 b^3 - 60 b^4 + 7 a b^5) + b^2(-12 + 7 a
b)(-16 + 7 b^4) \sin^2 \theta ) \sin 2\theta \nonu \\
&-& \left. 2 a^2 b (-96 - 46 a^4 + 6 a^5 b + 50 a^2 b^2 + 11 a^3 b^3 + 38 b^4
+ 2 a b (8- 21 b^4)) \sin 4\theta) \right], 
\nonu \\
R_{5}^{\,\,5} & = &  \frac{1}{48 L^2 \left[(a^2 + 7 b^2)^2 -112(a 
  b-1)\right]
(a^2  \cos^2 \theta + b^2  \sin^2
  \theta)^{\frac{8}{3}}} 
\nonu \\
& \times & \left[ 8 a^{14} \cos^4 \theta - 1120 a^{11} b \cos^4 \theta
  +
8 a^{12} b^2 \cos^2 \theta (15 + 13 \cos 2\theta) -112 a^9 b^3 \cos^2
\theta (83 \right.
\nonu \\
&+& 63 \cos 2\theta) - 8 a^5 b^3 ( 128 \cos^2 \theta (181 + 8 \cos
2\theta) + 7 b^4 (617 + 148\cos 2\theta -177 \cos 4\theta)) \nonu \\
&+ & 4 a^8 b^2 ( 4 \cos^2 \theta (1175 + 1091 \cos 2\theta) + 7 b^4 (
171 + 192 \cos 2\theta + 29 \cos 4\theta)) \nonu \\
&+& a^{10} ( 16 \cos^2 \theta (25 + 31 \cos 2\theta) + b^4 ( 941 +
1172 \cos 2\theta + 239 \cos 4\theta)) \nonu \\
&+ & a^6 ( 1024 \cos^2 \theta ( -7 + 3 \cos 2\theta) - 49 b^8 (-221
-172 \cos 2\theta + \cos 4\theta) \nonu \\
& + & 8 b^4 (8153 + 8832 \cos 2\theta +
879 \cos 4\theta))  
- 8 a^7 b ( 64 \cos^2 \theta (9 + 28 \cos 2\theta) \nonu \\
& + & 7 b^4 ( 564 \cos
2\theta + 73 ( 7 + \cos 4\theta)))
 -  16 a^3 b ( -1536 \cos^2 \theta + 49 b^8 ( 83 + \cos 2\theta) \nonu
 \\
& + & 32
 b^4 ( 747 + 380 \cos 2\theta)) \sin^2 \theta + 96 b^2 ( 768 + 1792
 b^4 + 245 b^8) \sin^4 \theta \nonu \\
& - & 96 a b^3 ( 2816 + 3584 b^4 + 343 b^8)
 \sin^4 \theta - 4 a^2 \sin^2 \theta ( 18432 \cos^2 \theta \nonu \\
& + &  256 b^4 (
 -323 + 15 \cos 2\theta) + 196 b^8 ( -181 + 115 \cos 2\theta) - 4802
 b^{12} \sin^2 \theta) \nonu \\
& - &  8 a^4 b^2 ( 64 ( -116 - 102 \cos 2\theta + 15 \cos 4\theta) +
b^4 ( -10065 + 1324 \cos 2\theta + 4625 \cos 4\theta) \nonu \\
& - & \left. 343 b^8 ( 9 + 5 \cos 2\theta ) \sin^2 \theta ) \right],
\nonu \\
R_{6}^{\,\,6} & = &  \frac{1}{48 L^2 \left[(a^2 + 7 b^2)^2 -112(a 
  b-1)\right]
(a^2  \cos^2 \theta + b^2  \sin^2
  \theta)^{\frac{8}{3}}} 
\nonu \\
& \times & \left[ 8 a^{14} \cos^4 \theta - 1120 a^{11} b \cos^4 \theta +
8 a^{12} b^2 \cos^2 \theta ( 15 + 13 \cos 2\theta ) - 112 a^9 b^3
\cos^2 \theta ( 83 \right. \nonu \\
& + & 63 \cos 2\theta) - 8 a^5 b^3 ( 128 \cos^2 \theta ( 100 + 89 \cos
2\theta) + 7 b^4 ( 617 + 148 \cos 2\theta - 177 \cos 4\theta)) \nonu
\\
& + & 4 a^8 b^2 ( 4\cos^2 \theta ( 1187 + 1079 \cos 2\theta) + 7 b^4 (
171 + 192 \cos 2\theta + 29 \cos 4\theta)) \nonu \\
&+& a^{10} ( 896 \cos^4 \theta + b^4 ( 941 + 1172 \cos 2\theta + 239
\cos 4\theta)) + a^6 ( - 128 \cos^2 \theta ( 17 + 15 \cos 2\theta)
\nonu \\
& - & 
49 b^8 ( -221 - 172 \cos 2\theta + \cos 4\theta) + 16 b^4 ( 3895 +
4440 \cos 2\theta + 597 \cos 4\theta ))
\nonu \\
& - & 8 a^7 b ( 64 \cos^2 \theta ( 18 + 19 \cos 2\theta) + 7 b^4 ( 564
\cos 2\theta + 73 ( 7 + \cos 4\theta))) \nonu \\
& - & 16 a^3 b ( 768 \cos^2 \theta + 49 b^8 ( 83 + \cos 2\theta ) + 32
b^4 ( 216 + 281 \cos 2\theta )) \sin^2 \theta \nonu \\
& +& 96b^2 (-384 + 392 b^4 + 245 b^8) \sin^4 \theta - 96 a b^3 ( -1408
+ 896 b^4 + 343 b^8) \sin^4 \theta \nonu \\
& + & 8 a^2 \sin^2 \theta ( 4608 \cos^2 \theta
-  112 b^8 ( -92 + 29 \cos 2\theta) + 112 b^4 (-17+ 147 \cos 2\theta)
\nonu \\ 
& + &
2401 b^{12} \sin^2 \theta) 
 +  8 a^4 b^2 (4076 + 6480 \cos 2\theta + b^4 ( 6927 + 116 \cos 2\theta -
2927 \cos 4\theta ) \nonu \\
&+& \left.  2436 \cos 4\theta + 343 b^8 ( 9+ 5 \cos 2\theta) \sin^2
  \theta)
\right] =  R_{7}^{\,\, 7} = R_{8}^{\,\, 8} = R_{9}^{\,\,9} =R_{10}^{\,\,10}
=
R_{11}^{\,\, 11},
\nonu \\
%
R_{5}^{\,\,4} & = &  \frac{1}{8 L \sqrt{a} \sqrt{(a^2 + 7 b^2)^2 -112(a 
  b-1)}
(a^2  \cos^2 \theta + b^2  \sin^2
  \theta)^{\frac{8}{3}}} 
\nonu \\
& \times & \left[-2 (a^5(80 + a^4) - 96 a^2 (4 + a^4) b + 2 a^3(192+ 7
  a^4) b^2 -12 (16+9a^4) b^3 \right. \nonu \\ 
&+& 16 a ( 7 + 3 a^4) b^4 + 120 a^2 b^5 - 14 a^3 b^6 + 84 b^7 -49 a
b^8) \sin 2\theta \
\nonu \\
&- & (a-b)(a+b)( a^7 -192 b -92 a^4 b + 13 a^5 b^2 + 8 a^2 b^3 + 84
b^5 + 7 a b^2 (16- 7b^4)  \nonu \\ 
& + & \left. 5 a^3 (16 + 7 b^4)) \sin 4\theta \right].
\nonu 
\eea
The Ricci tensor before imposing the RG flow equations (\ref{flow}) to two vacuum
expectation values $(a,b)$ was given in \cite{AI}. The above Ricci
tensor holds for both spherical parametrizations in section 2.

\section{The coefficient functions for $G_2$ invariant flow }

The coefficient functions appearing in (\ref{fst1}) or (\ref{fst2}) are given by
\bea
\tilde{h}_1 & = & \frac{L^3 \, \sqrt{a b-1} \, \sin^4 \theta 
\left[ -a^4 \, \cos^2 \theta + a^2 \, b^2 (3 + 2 \, \cos 2\theta ) + 3 \,
b^4 \, \sin^2 \theta  \right]}
{2\, a\, (a^2 \, \cos^2 \theta + b^2 \, \sin^2 \theta)^2}, \nonu \\
h_2 & = & \frac{L^3 \, b \, \sqrt{a b -1} \, \sin^4 \theta}
{2 (a^2 \, \cos^2 \theta + b^2 \, \sin^2 \theta)}, \nonu \\
\tilde{h}_2 & = & \frac{L^3 \, b \, \sqrt{a b -1} \, \cos \theta
  \, \sin^3 \theta \, \left[ a^2 \, (3 + \cos 2\theta)+ 2 b^2\, 
\sin^2 \theta \right]}{2 
(a^2 \, \cos^2 \theta + b^2 \, \sin^2 \theta)^2}, \nonu \\
\widetilde{h}_3 & = & \frac{L^2 \, a^{\frac{3}{2}} \, \sqrt{a b-1} \,
\cos \theta \, \sin^3 \theta }
{2 \, \sqrt{(a^2 + 7 b^2)^2 -112(a b-1)} \,
(a^2 \, \cos^2 \theta + b^2 \, \sin^2 \theta)^2} \nonu \\
 & \times &  \left[-16 a^3  \cos^2 \theta + a^4  b
 (3 + \cos 2\theta) + 2  a^2 b^3 (11 + 3 \cos 2\theta) -112 
a  b^2 \sin^2 \theta \right. \nonu \\
& + & \left.  2  b (48 + 7 b^4)  \sin^2 \theta \right],
\nonu \\
\tilde{h}_4 & = & \frac{L^2 \, a^{\frac{1}{2}} \, \sqrt{a b-1} \,
\sin^4 \theta}{4 \, \sqrt{(a^2 + 7 b^2)^2 -112(a b-1)} \,
(a^2 \, \cos^2 \theta + b^2 \, \sin^2 \theta)^2}
 \left[2 a^6 \cos^2 \theta -64 a^3 b \cos^2 \theta  \right. \nonu \\
& + & 
a^4 b^2 (1 + 3 \cos 2\theta) + a^2(96 \cos^2 \theta - 5 b^4(9 + 5 \cos
2\theta))+
128 a b^3 \sin^2 \theta \nonu \\
& - & \left. 6 b^2(16+ 7b^4)\sin^2 \theta \right], \nonu \\
\tilde{h}_5 & = & \frac{2 \,L^2 \, \sqrt{a b -1} \, (7 b^3 + a^2 b
  - 4 a) \, \sin^2 \theta  }
{\sqrt{a} \, \sqrt{(a^2 + 7 b^2)^2 -112(a b-1)}}, \nonu \\
\nonu \\
W_r & = &  -\frac{1}{2L}\,a^2 \left[
a^5 \, \cos^2 \theta +a^2 \, b\,(ab-2)\,(4+3\cos 2\theta)
+b^3 \,(7ab-12) \, \sin^2 \theta \right], \nonu \\
W_{\theta} & = &  -\frac{a^{\frac{3}{2}} \,
\left[48\,(1-ab)+(a^2 -b^2)\,(a^2 +7b^2)\right]}
{
\dsqrt{(a^2 +7b^2)^2 -112\,(ab-1)}}\,\sin \theta \, \cos \theta. 
\nonu 
\eea

\section{The seven frames 
on ${\bf S}^6$ with ${\bf CP}^2$ space: $G_2$-invariant flow }

The seven frames ($\a_i \equiv 2 \th_i, \, i =1, 2, 3$) in subsection
2.2 are given by
\bea
\hat{\bf e}_1^1  & = & \frac{1}{2} \cos (\a_2 +\a_3
+\theta_5) \sin \a_1 \sin \theta_4 \sin \theta_6,
\nonu \\
\hat{\bf e}_1^2  & = & \frac{1}{2} \sin \a_1 \sin \th_4 \sin
(\a_2+\a_3 +\th_5) \sin \th_6, \nonu \\
\hat{\bf e}_1^3  & = & -\frac{1}{2} \cos \a_1 \cos
(-\a_2 +\a_3 +\th_5) \sin \th_4 \sin \th_6, 
\nonu \\
\hat{\bf e}_1^4  & = & -\frac{1}{2} \cos \a_1 \sin \th_4 \sin
(-\a_2+\a_3
+ \th_5) \sin \th_6, 
\nonu \\
\hat{\bf e}_2^1  & = & \frac{1}{2} \cos \a_1 \sin \th_4 
\sin (\a_2 +\a_3
+\theta_5)  \sin \theta_6,
\nonu \\
\hat{\bf e}_2^2  & = &  -\frac{1}{2} \cos \a_1 \cos
(\a_2+\a_3 +\th_5) \sin \th_4 \sin \th_6,
\nonu \\
\hat{\bf e}_2^3  & = & -\frac{1}{2} \sin \a_1 
\sin \th_4  \sin
(-\a_2 +\a_3 +\th_5)  \sin \th_6, \nonu \\
\hat{\bf e}_2^4  & = & \frac{1}{2}  \cos
(-\a_2+\a_3
+ \th_5) \sin \a_1 \sin \th_4 \sin \th_6, 
\nonu \\
\hat{\bf e}_3^1  & = & \hat{\bf e}_2^1, 
\qquad
\hat{\bf e}_3^2 = \hat{\bf e}_2^2,
\qquad
\hat{\bf e}_3^3   =  -\hat{\bf e}_2^3,
\qquad
\hat{\bf e}_3^4 = -\hat{\bf e}_2^4,
\nonu \\
\hat{\bf e}_4^1  & = & -\frac{1}{2} \cos \a_1 \cos \th_4 
\cos (\a_2 +\a_3
+\theta_5)  \sin \theta_6,
\nonu \\
\hat{\bf e}_4^2  & = & -\frac{1}{2} \cos \a_1 \cos \th_4 \sin
(\a_2+\a_3 +\th_5)  \sin \th_6,
\nonu \\
\hat{\bf e}_4^3  & = & - \cos \th_4  
 \cos
(-\a_2 +\a_3 +\th_5)  \sin \a_1 \sin \th_6, 
\nonu \\
\hat{\bf e}_4^4  & = & - \cos \th_4 \sin \a_1 \sin
(-\a_2+\a_3
+ \th_5)  \sin \th_6, 
\nonu \\
\hat{\bf e}_4^5 & = & \cos \th_5 \sin \th_4 \sin \th_6, \qquad
\hat{\bf e}_4^6 = \sin \th_4 \sin \th_5 \sin \th_6,   
\nonu \\
\hat{\bf e}_5^1  & = &  2 \hat{\bf e}_2^1,
\qquad
\hat{\bf e}_5^2 = 
 2 \hat{\bf e}_2^2,
\qquad
\hat{\bf e}_5^3   =   -  2 \hat{\bf e}_2^3,
\qquad
\hat{\bf e}_5^4 = -   2 \hat{\bf e}_2^4,
\nonu \\
\hat{\bf e}_5^5 & = & \cos \th_4 \sin \th_5 \sin \th_6, \qquad
\hat{\bf e}_5^6 = -\cos \th_4 \cos \th_5 \sin \th_6,   
\nonu \\
\hat{\bf e}_6^1  & = &  -\cos \a_1  
\cos (\a_2 +\a_3
+\theta_5)  \cos \theta_6 \sin \th_4,
\nonu \\
\hat{\bf e}_6^2  & = & - \cos \a_1 \cos \th_6 \sin \th_4 \sin 
(\a_2+\a_3 +\th_5),
\nonu \\
\hat{\bf e}_6^3  & = &  -  
 \cos
(-\a_2 +\a_3 +\th_5)  \cos \th_6 \sin \a_1 \sin
\th_4, \nonu \\
\hat{\bf e}_6^4  & = & -  \cos \th_6 \sin \a_1 \sin \th_4 \sin
(-\a_2+\a_3
+ \th_5), 
\nonu \\
\hat{\bf e}_6^5 & = & -\cos \th_4 \cos \th_5 \cos \th_6, \qquad
\hat{\bf e}_6^6 = -\cos \th_4 \cos \th_6 \sin \th_5,   
\qquad
\hat{\bf e}_6^7  =  \sin \th_6,
\nonu \\
\hat{\bf n}^1  & = &  -2   \hat{\bf e}_2^2,
\qquad
\hat{\bf n}^2 = 2   \hat{\bf e}_2^1,
\qquad
\hat{\bf n}^3   =    2 \hat{\bf e}_2^4,
\nonu \\
\hat{\bf n}^4  & = &  -  2 \hat{\bf e}_2^3,
\qquad
\hat{\bf n}^5  =  -  \hat{\bf e}_5^6, 
\qquad
\hat{\bf n}^6 =  \hat{\bf e}_5^5,
\qquad
\hat{\bf n}^7  =  \cos \th_6.
\nonu
\eea

\section{The $\hat{F}_{ij}, \hat{T}_{ijk}$ and $\hat{S}_{ijk}$ tensors
on ${\bf S}^6$ with ${\bf CP}^2$ space: $G_2$-invariant flow }

The three invariant tensors appearing in (\ref{fst2})
are given as follows.
The $\hat{F}_{ij}$ tensor appearing in the 4-forms $F_{45mn}$ where 
$\a_i \equiv 2 \th_i$($i =1, 2, 3$) for simplicity is given by
\bea
\hat{F}_{12}  & = & \frac{1}{32} \sin^2 \theta_4 \sin^2 \theta_6
\left[ -4 \cos 2\alpha_2 \cos \theta_6 + 4 \cos 2\alpha_1 \cos
  2(\alpha_3 +\theta_5) \cos \theta_6 \right. \nonu \\
& + & ( 8 \cos \alpha_1 \cos
  (\alpha_2 - \alpha_3-\theta_5) \sin \theta_4 -\cos \theta_4  ( \cos 
(2\alpha_1-2\alpha_3-3\theta_5)\nonu \\
& - &  \cos (2\alpha_1-2\alpha_3 -\theta_5)-
\cos (2\alpha_1 + 2\alpha_3+\theta_5) + \cos (2\alpha_1 +
2\alpha_3+3 \theta_5) \nonu \\
& + &  \left. 8 \cos \alpha_1 \cos \theta_5 \sin \alpha_1 +
8 \cos \alpha_2 \sin \alpha_2 \sin \theta_5)) \sin \theta_6
\right],
 \nonu \\
\hat{F}_{13}  & = & \frac{1}{32} \sin^2 \theta_4 \sin^2 \theta_6 
\left[ -2( \cos 2(\alpha_1 -\alpha_2) + \cos 2(\alpha_1 + 
\alpha_2) -2 \cos 2(\alpha_3 + \theta_5)) \cos \theta_6 
\right. \nonu \\
& + & ( 8 \cos \alpha_1 \cos (\alpha_2-\alpha_3-\theta_5) 
\sin \theta_4 + \cos \theta_4 ( \cos (2\alpha_1-2\alpha_2-
\theta_5) \nonu \\
& - & \cos (2\alpha_1 +2\alpha_2-\theta_5)
-\cos (2 \alpha_1 -2\alpha_2 +\theta_5) + \cos 
(2\alpha_1 +2\alpha_2 +\theta_5) 
\nonu \\
&+& \left. 8  \cos \alpha_1 \cos \theta_5 \sin \alpha_1 + 4 \sin \theta_5
\sin 2(\alpha_3 + \theta_5))) \sin \theta_6 \right],
\nonu \\
\hat{F}_{14}  & = & \frac{1}{4}
\sin \theta_4 \sin^2 \theta_6 \left[
\cos \theta_4 \cos \theta_6 ( - \sin 2\alpha_2 + \sin
2(\alpha_3 +\theta_5) + 2 \cos \theta_6 \sin 
\theta_4 ( \right. \nonu \\
&+& \cos \alpha_1 \cos (\alpha_2 -\alpha_3-\theta_5) \sin 
\theta_5 - \cos \theta_5 \sin \alpha_1 \sin (\alpha_2 +
\alpha_3 +\theta_5)) \nonu \\
& + & \left.
(\cos 2\alpha_2 -\cos 2(\alpha_3 +\theta_5)) \sin \theta_5
\sin \theta_6 \right], 
\nonu \\
\hat{F}_{15}  & = & \frac{1}{8} \sin \theta_4 
\sin^2 \theta_6 \left[ -4 \cos \theta_4 \cos 
\theta_6 \sin \alpha_1 \sin \theta_5 \sin (\alpha_2 +
\alpha_3 + \theta_5) \right.
\nonu \\
& + & 4 \cos^2 \theta_4 \cos (\alpha_2 +\alpha_3 +
\theta_5) \sin \alpha_1 \sin \theta_6 + 4 \cos^3 
\alpha_1 \cos (\alpha_2 -\alpha_3 -\theta_5) 
\sin^2 \theta_4 \sin \theta_6 \nonu \\
& + & \sin 2 \theta_4 ( \cos \theta_5 \sin 2 \alpha_1
+ \sin^2 \alpha_1 \sin \theta_5 ( \sin 2\alpha_2 +
\sin 2(\alpha_3 +\theta_5))) \sin \theta_6 \nonu \\
&+& 2\cos 2\alpha_2 \sin^2 \alpha_1 \sin \theta_4 (
\cos \theta_6 - \cos \theta_4 \cos \theta_5 
\sin \theta_6 ) \nonu \\
& + & 2 \cos 2(\alpha_3 +\theta_5) 
\sin^2 \alpha_1 \sin \theta_4 (   \cos \theta_6 + \cos \theta_4 \cos
\theta_5 \sin \theta_6 ) \nonu \\
& + & 4 \cos \a_1 \cos (\a_2-\a_3-\th_5) ( -\cos \th_4 \cos \th_5 \cos
\th_6 + \sin^2 \a_1 \sin^2 \th_4 \sin \th_6) \nonu \\
&+&  2 \cos^2 \a_1 (( -\cos 2\a_2 + \cos 2(\a_3 +\th_5) \cos \th_6 \sin
\th_4 \nonu \\
&+ & \left.  \sin (\a_2 +\a_3) \sin 2\th_4 \sin (\a_2 -\a_3 -\th_5) \sin
\th_6) \right], 
\nonu \\
\hat{F}_{16}  & = & \frac{1}{4} \sin \th_4 \sin \th_6 
\left[ -2 \cos \a_1 \cos \th_4 \cos (\a_2-\a_3-\th_5) \sin \th_5 
\right. \nonu \\ 
& + & \left. \sin
  \th_4 ( - \sin 2\a_2 + \sin 2(\a_3 +\th_5)) 
+  2 \cos \th_4 \cos \th_5 \sin \a_1 \sin (\a_2 +\a_3 +\th_5) 
\right], \nonu \\
\hat{F}_{23}  & = & \frac{1}{4} \cos \a_1 \sin \a_1 \sin^2 \th_4
\sin^2 \th_6 \left[ \cos \th_6 (\sin 2\a_2 + \sin 2(\a_3 +\th_5)) 
\right. \nonu \\
&-& \left. ( 2\cos \a_1 \sin \th_4 \sin (\a_2-\a_3 -\th_5) +\cos \th_4
  (
\cos 2\a_2 + \cos 2(\a_3 +\th_5)) \sin \th_5 ) \sin \th_6 \right],  
\nonu \\
%
\hat{F}_{24}  & = & \frac{1}{8} \sin \th_4 \sin^2 \th_6 \left[
-2 \cos \th_4 \cos 2(\a_3 +\th_5) \cos \th_6 \sin 2\a_1 \right.
\nonu \\
&+& \cos^2 \th_4 ( 4 \cos \th_5 ( \sin^2 \a_1 - \cos 2\a_3 \sin 2\a_1
\sin^2 \th_5 ) \nonu \\
&+ & \sin 2\a_1 \sin 2\a_3 ( \sin \th_5 - \sin 3 \th_5)) \sin \th_6
\nonu \\ 
& + &
2\sin \th_4 ( 2\cos \a_1 \cos \th_5 \cos (\a_2+\a_3+\th_5) \cos \th_6
- 
2 \cos \th_6 \sin \a_1 \sin (\a_2-\a_3-\th_5) \sin \th_5 \nonu \\
& + & \left. 
\sin \th_4 ( 2 \cos \th_5 \sin^2 \a_1-\sin 2\a_1 \sin \th_5 \sin
2(\a_3 +\th_5)) \sin \th_6)
\right],
\nonu \\
\hat{F}_{25}  & = &  \frac{1}{8} \sin \th_4 \sin^2 \th_6 \left[
 2 \cos \th_6 ( 2 \cos \th_4 ( \cos \th_5 \sin \a_1 \sin (\a_2 -\a_3
 -\th_5) \right. \nonu \\ 
&+&  \cos \a_1 \cos (\a_2 +\a_3 +\th_5) \sin \th_5 ) + \sin 2\a_1 \sin
\th_4 ( \sin 2\a_2 + \sin 2(\a_3 +\th_5)) ) \nonu \\
& + & ( -8 \cos^2 \a_1 \sin \a_1
\sin^2 \th_4 \sin (\a_2-\a_3-\th_5) +
4 \cos \a_1 \cos^2 \th_4 \sin (\a_2+\a_3 +\th_5) 
\nonu \\
&+& \left. 
\sin 2\th_4 ( 2\sin^2 \a_1 \sin \th_5 + \sin 2\a_1 ( - \cos 2\a_2
\sin \th_5 + \sin (2\a_3 +\th_5)) ) ) \sin \th_6 \right],
\nonu \\  
\hat{F}_{26}  & = & \frac{1}{16} \sin \th_4 \sin \th_6 \left[
(2-3 \cos 2(\a_3+\th_5)) \sin 2\a_1 \sin \th_4 \right.
\nonu \\
& - & 2 \cos \a_1 ( 
4 \cos \th_4 \cos \th_5 \cos (\a_2 +\a_3 +\th_5) + ( 2 + \cos 2(\a_3
+\th_5)
\sin \a_1 \sin \th_4 )  
\nonu \\
&+& 8 \left.
\cos \th_4 \sin \a_1 \sin (\a_2 -\a_3 -\th_5) \sin \th_5 \right],
\nonu \\
\hat{F}_{34}  & = & \frac{1}{4} \sin \th_4 \sin^2 \th_6 \left[
\cos \th_6 ( \cos^2 \a_2 \cos \th_4 \sin 2\a_1 -\cos \th_4 \sin 2\a_1
\sin^2 \a_2 \right. 
\nonu \\
& + & 2\cos \a_1 \cos \a_2 \cos \th_5 \cos (\a_3 +\th_5) \sin \th_4 +
2 \cos (\a_3 +\th_5) \sin \a_1 \sin \a_2 \sin \th_4 \sin \th_5 
   \nonu \\
& - & 2 \sin \th_4 ( \cos \a_1 \cos \th_5 \sin \a_2 + \cos \a_2 \sin
\a_1 \sin \th_5 ) \sin (\a_3 +\th_5) ) \nonu \\
&+& \left. 
(- 2\cos \th_5 \sin^2 \a_1 + \sin 2\a_1 \sin 2\a_2 \sin \th_5) \sin
\th_6 \right], \nonu \\
\hat{F}_{35}  & = & \frac{1}{2} \cos \th_4 \sin \th_4 \sin^2 \th_6
\left[ - \sin^2 \a_1 \sin \th_4 \sin \th_5 \sin \th_6 \right.
\nonu \\
&-& \cos \th_5 \sin \a_1 ( \cos \th_6 \sin (\a_2-\a_3-\th_5) + \cos
\a_1 \sin 2\a_2 \sin \th_4 \sin \th_6 ) \nonu \\ 
& + & \left. \cos \a_1 ( \cos (\a_2 +\a_3 +\th_5) \cos \th_6 \sin \th_5 +
\cos \th_4 \sin (\a_2 +\a_3 +\th_5) \sin \th_6 ) 
\right], 
\nonu \\
\hat{F}_{36}  & = & \frac{1}{4} \sin \th_4 \sin \th_6 \left[ \cos^2 \a_2 \sin
  2\a_1 \sin \th_4 - 2 \cos \a_1 ( \cos \th_4 \cos \th_5 \cos (\a_2
  +\a_3 +\th_5) \right. \nonu \\
& + & \left. \sin \a_1 \sin^2 \a_2 \sin \th_4 ) - 2 \cos \th_4 \sin \a_1 \sin
(\a_2 -\a_3-\th_5) \sin \th_5 \right],
\nonu \\
\hat{F}_{45}  & = & -\frac{1}{4} \sin^2 \th_6 \left[
\cos \th_6 ( 2\cos \a_1 ( - \cos(\a_2 +\a_3) \cos 2\th_4 + \cos (\a_2
+\a_3 +2\th_5) ) \right. \nonu \\
&+ & 2 ( \cos (\a_2 -\a_3) \cos 2\th_4 + \cos (\a_2 -\a_3 -2\th_5) )
\sin \a_1 + \cos 2\a_2 \sin 2\a_1 \sin 2\th_4 ) \nonu \\
&+&  4( - \cos \th_5 \sin^2 \a_1 \sin \th_4 + \cos \a_1 ( \cos \th_4
\cos (\a_2 +\a_3 +\th_5) \nonu \\
& + & \left. \sin \a_1 \sin 2\a_2 \sin \th_4 \sin \th_5) )
\sin \th_6 \right],
\nonu \\
\hat{F}_{46}  & = & \frac{1}{4} \sin \th_6 \left[ 
\cos (\a_1+\a_2 -\a_3) - \cos (\a_1-\a_2 +\a_3)-\cos(\a_1+\a_2
-\a_3-2\th_5) \right. \nonu \\
& + & \left. \cos (\a_1 -\a_2 +\a_3 +2 \th_5) 
 - 4 \cos \a_1 \cos \th_5 \sin (\a_2 + \a_3 +\th_5) \right],    
\nonu \\
\hat{F}_{56}  & = & \sin \th_4 \sin \th_6
\left[ -\cos \a_1 \cos (\a_2 +\a_3) \cos \th_4 + \cos(\a_2 -\a_3) \cos
  \th_4 \sin \a_1 \right. \nonu \\
&+& \left. \frac{1}{2} \cos 2\a_2 \sin 2\a_1 \sin \th_4 \right]. 
\nonu
\eea
The $\hat{F}^{ij}$ tensor appearing in the 4-forms $F_{mnpq}$ is given by
\bea
\hat{F}^{12}  & = & -2 \csc \th_4 \csc \th_6 \left[
\cos 2\a_2 \cot \th_6 \csc \th_4 + \cos \th_5 (
\cot \a_1 \cot \th_4 -\cos (\a_2-\a_3) \sec \a_1 ) \right.
\nonu \\
& + & \left.
( - \cos \a_3 \sec \a_1 \sin \a_2 + \cos \a_2 ( 2 \cot \th_4
\sin \a_2 + \sec \a_1 \sin \a_3 ) ) \sin \th_5 
\right],    \nonu \\ 
\hat{F}^{13}  & = & 2 \csc^2 \th_6 \left[
\cos \th_6 ( \cos (2\a_3 +2\th_5) \csc^2 \th_4 + \csc 2\th_4
( \cos (\a_1 +\a_2 -\a_3) + \cos (\a_1 -\a_2 +\a_3) \right.
   \nonu \\
&+ &  \cos (\a_1 +\a_2 -\a_3-2\th_5) + \cos (\a_1-\a_2 +\a_3 +
2\th_5 ) + 4 \sin \a_1 \sin \th_5 \sin (\a_2 +\a_3 +\th_5) ) ) 
\nonu \\
&+ & ( \cos \th_5 ( \cot \a_1 \cot \th_4 \csc \th_4 + ( \cos 2\a_2 
- \cos (2\a_3 +2\th_5) ) \sec \th_4 ) \nonu \\
& + & 
\csc \th_4 ( \cos (\a_2 -\a_3 -\th_5) \sec \a_1 -2 
\cos (\a_2 +\a_3 + \th_5) \sin \a_1 \nonu \\
& + & \left. \cot \th_4 \sin \th_5 \sin 
(2\a_3 +2\th_5) ) ) \sin \th_6 \right],  
\nonu \\ 
\hat{F}^{14}  & = & \csc \th_6 \left[
2\cos \a_1 \cos (\a_2 -\a_3 -\th_5) \cot \th_6 \sin \th_5 + (
\cos 2\a_2 - \cos (2\a_3 +2\th_5) ) \csc \th_4 \sin \th_5 \right.
\nonu \\
& + & \left. \cot \th_4 \cot \th_6 (-\sin 2\a_2 + \sin (2\a_3 +2\th_5))
-2 \cos \th_5 \cot \th_6 \sin \a_1 \sin (\a_2 +\a_3 +\th_5) 
\right],  \nonu \\ 
\hat{F}^{15}  & = & \frac{1}{4}  \csc \th_6 \sec \th_4 \left[
-\cos (2\a_1 -2\a_2 -\th_5) + \cos (2\a_1 +2\a_2 -\th_5)  + 
\cos (2\a_1-2\a_2 +\th_5) \right. \nonu \\
& - & \cos (2\a_1 + 2\a_2 +\th_5) -
2 ( \cos (2\a_2 -\th_5) + \cos (2\a_2 +\th_5) -\cos (2\a_3 +\th_5)
- \cos (2\a_3 + 3\th_5) \nonu \\
& - &  4 \cos (\a_2 +\a_3 +\th_5) \cot \th_4 \sin \a_1  
+ 2 \cos 2\a_1 \sin 2\a_2 \sin \th_5 + \cot \th_6 \csc 
\th_4 ( \cos (\a_1 +\a_2-\a_3) \nonu \\   
& + & \cos (\a_1 -\a_2 +\a_3) + \cos (\a_1 +\a_2 -\a_3 -
2\th_5) + \cos (\a_1 -\a_2 +\a_3 +2\th_5) \nonu \\
&+ &  \left. 4
\sin \a_1 \sin \th_5 \sin (\a_2 +\a_3 +\th_5) ) ) \right], 
\nonu \\ 
\hat{F}^{16}  & = & \csc \th_6 \left [
-\sin 2\a_2 - 2 \cos \a_1 \cos (\a_2 -\a_3-\th_5) \cot 
\th_4  \sin \th_5 \right.
\nonu \\
&+& \left.
\sin (2\a_3 +2\th_5) + 2 \cos \th_5 \cot \th_4 \sin \a_1 
\sin (\a_2 +\a_3 +\th_5) \right], \nonu \\ 
\hat{F}^{23}  & = & \csc 2\a_1 \csc^2 \th_6 \left[
2 \cos \th_6  ( - 4 \csc 2\th_4 ( \cos \a_1 \cos 
\th_5 \sin (\a_2 -\a_3 -\th_5) \right. \nonu \\
& +  & \cos (\a_2 +\a_3 +\th_5) \sin \a_1 \sin \th_5 )
+ \csc^2 \th_4 ( \sin 2\a_2 + \sin (2\a_3 +2\th_5) ) ) \nonu \\
&- & 2 ( \csc \th_4 ( 2 \cos \a_1 \sin (\a_2 -\a_3 -\th_5)
+ \cos 2\a_2 \cot \th_4 \sin \th_5  ) + \sec \th_4 
( \cos \th_5 ( \cos 2\a_1 \sin 2\a_2 \nonu \\
& + & \sin 2\a_3 ) + ( \cos 2\a_3 + \cos (2\a_3 +2\th_5) \csc^2 
\th_4 + \sin 2\a_1 ) \sin \th_5 ) \nonu \\
&+ &  \left.
2\csc \th_4 \sin \a_1 \sin (\a_2 +\a_3 +\th_5) ) \sin \th_6
\right],      \nonu \\ 
\hat{F}^{24}  & = & -\csc 2\a_1 \csc \th_4 \csc^2 \th_6
\left[ \cos \th_4 \cos (2\a_3 +2\th_5) \cos \th_6 +
\cos^2 \th_4 \sin \th_6 ( \sin \th_5 \sin (2\a_3 +2\th_5)
\right.  \nonu \\ 
&-& \cos \th_5 \tan \a_1 ) + \cos 2\a_1 ( \cos^2 \a_2 \cos \th_4
\cos \th_6 - \cos \th_4 \cos \th_6 \sin^2 \a_2  \nonu \\
&+ & \cos (\a_3 +\th_5) \cos \th_6 \sec \a_1 \sin \a_2 \sin 
\th_4 \sin \th_5 \nonu \\
& + & \cos \a_2 \cos \th_6 \sin \th_4 
( \cos \th_5 \cos (\a_3 +\th_5) \csc \a_1  -
\sec \a_1 \sin \th_5 \sin (\a_3 +\th_5) ) \nonu \\
&+ & \sin 2\a_2 \sin \th_5 \sin \th_6 
-\cos \th_5 ( \cos \th_6 \csc \a_1 \sin \a_2 \sin \th_4 
\sin (\a_3 +\th_5) + \sin \th_6 \tan \a_1 ) ) \nonu \\
& + & \sin \th_4 ( \sin \th_5 (
\cos \th_6 \sec \a_1 \sin (\a_2 -\a_3 -\th_5) + 
\sin \th_4 \sin (2\a_3 +2\th_5) \sin \th_6 ) \nonu \\
&-& \left.
\cos \th_5 (\cos (\a_2 +\a_3 +\th_5) \cos \th_6 \csc \a_1 +
\sin \th_4 \sin \th_6 \tan \a_1 ) ) ) \right],  \nonu \\ 
\hat{F}^{26}  & = & - \csc \th_6 \left[
\cos 2\a_2 \cot 2\a_1 + \csc 2\a_1  (
\cos (2\a_3 +2\th_5) \right. \nonu \\
& + & \left. 
2 \cot \th_4 ( \cos \th_5 \cos (\a_2 +\a_3 +
\th_5) \sin \a_1 
 -  \cos \a_1 \sin (\a_2 -\a_3 -\th_5) \sin \th_5 ) ) 
\right],     
\nonu \\
%
\hat{F}^{25}  & = & \csc 2\a_1 \csc^2 \th_6 \sec \th_4 
\left[ 2 \cos \th_6 \csc \th_4 ( \cos \a_1 \cos \th_5 \sin 
(\a_2 -\a_3-\th_5) \right. \nonu \\ 
& + & \cos (\a_2 +\a_3 +\th_5) \sin \a_1 \sin \th_5 ) +
( \cos 2\a_1 \cos \th_5 \sin 2\a_2 + ( \cos (2\a_3 +2\th_5) +
\sin 2\a_1 ) \sin \th_5 \nonu \\
& + & \left. 2 \cot \th_4 \sin \a_1 \sin (\a_2 +\a_3 +\th_5)
+ \sin (2\a_3+\th_5) ) \sin \th_6 \right], \nonu \\
\hat{F}^{34}  & = & \csc \th_6 \left[
\cos (2\a_3 +2\th_5) \cot 2\a_1 \cot \th_4 \cot \th_6 +
\cos 2\a_2 \cot \th_4 \cot \th_6 \csc 2\a_1 \right.
\nonu \\
& - & 2 \cos \a_1 \cos (\a_2 +\a_3 +\th_5) \sec \th_4 -
\cos \th_5 ( \csc \th_4 + 2 \cot \th_6 \sin \a_1 ( 
 \nonu \\
& & -\cos \a_2 \cos (\a_3 +\th_5) (-1 + \csc 2\a_1) + (1 +\csc 
2\a_1 ) \sin \a_2 \sin (\a_3 + \th_5) ) ) \nonu \\
&+ & \sin \th_5 ( \csc 2\a_1 \csc \th_4 \sin 2\a_2 + \cot 2\a_1 
\csc \th_4 \sin (2\a_3 + 2\th_5) \nonu \\
& + & \left.
\cot \th_6 ( \csc \a_1 \sin (\a_2 -\a_3 -\th_5) +
2 \cos \a_1 \sin (\a_2 +\a_3 +\th_5) ) ) 
\right],  \nonu \\ 
\hat{F}^{35}  & = & -\frac{1}{2} \csc \th_6 
\left[ 2 \cos \th_5 \sec \th_4 ( \csc 2\a_1 \sin 2\a_2 +
\cot \th_6 \csc \a_1 \csc \th_4 \sin (\a_2 -\a_3 -\th_5) )
\right. \nonu \\
& + & \cos 2\a_1  \csc^2 \a_1 \csc \th_4 \sec \a_1 \sin (\a_2 +
\a_3 +\th_5 ) -\csc^2 \a_1 \csc \th_4 \sec \a_1 \sec^2 \th_4 \sin 
(\a_2 +\a_3 +\th_5) \nonu \\   
& + & \sec \th_4 ( 2 ( 1+ \cos (2\a_3 +2\th_5) \cot 2\a_1 -
\cos (\a_2 +\a_3 +\th_5) \cot \th_6 \csc \th_4 \sec \a_1 )
\sin \th_5  \nonu \\
& + & \left. 
2 \cot 2\a_1 \sin (2\a_3 +\th_5) +\csc^2 \a_1 \sec \a_1 ) 
\sin (\a_2 +\a_3 + \th_5) \tan \th_4 ) \right],   
\nonu \\ 
\hat{F}^{36}  & = & \frac{1}{4} \csc \th_6 
\left[ \cot 2\a_1 ( -2 + 4 \cos (2\a_3 +2\th_5) +
4\cos \a_1 \csc 2\a_1 ( 2\cos \th_5 \cos (\a_2 +\a_3 +\th_5)
\cot \th_4 \right.  \nonu \\
&+ & \sin \a_1 ) - 4 \cot \th_4 \sec \a_1 \sin (\a_2 -\a_3-\th_5)
\sin \th_5 ) + 4\csc^2 2\a_1 ( -\cos^2 2\a_1 + \csc^2 \th_4 ) 
\sec \th_4 \nonu \\
& & ( \cos^2 \a_2 \sin 2\a_1 \sin \th_4 - 2\cos \a_1 ( \cos 
\th_4 \cos \th_5 \cos (\a_2 +\a_3 +\th_5) \nonu \\
& + & \sin \a_1 \sin^2 \a_2 \sin \th_4 ) - 2 \cos \th_4 
\sin \a_1 \sin (\a_2 -\a_3 -\th_5) \sin \th_5  ) \tan \th_4 \nonu \\
& -& 4 \left.
\tan \th_4 ( - 2 \cos \a_1 \cos (\a_2 +\a_3) +
2\cos (\a_2 -\a_3) \sin \a_1 + \cos 2\a_2 \sin 2\a_1 \tan \th_4 )
\right],    \nonu \\    
\hat{F}^{45}  & = & -\frac{1}{2} \csc \th_6 
\left[ 2 \cos (\a_2 -\a_3 -\th_5) \cos \th_5 \cot \th_6 \sin 
\a_1 + \cos \a_1 ( (- \cos (\a_2 +\a_3 ) \cos 2\th_4 \right.
\nonu \\ 
& + & \left.
\cos (\a_2 +\a_3 +2\th_5) ) \cot \th_6 \sec^2 \th_4 +
2 \cos (\a_2 +\a_3 +\th_5) ( \sec \th_4 -\cos \th_5 \cot \th_6 
\tan^2 \th_4 ) ) \right],   
\nonu \\ 
\hat{F}^{46}  & = & \frac{1}{4} \csc \th_6 \left[
\cos (\a_1 +\a_2 -\a_3) -\cos (\a_1 -\a_2 +\a_3) -\cos (\a_1 +
\a_2 -\a_3 -2\th_5) \right. \nonu \\
 & + & \left.
\cos (\a_1-\a_2 +\a_3 +2\th_5) - 4 \cos \a_1 \cos \th_5
\sin (\a_2 +\a_3 +\th_5) \right],  \nonu \\ 
\hat{F}^{56}  & = & \frac{1}{4} \csc \th_6 \left[ -\cos (\a_1 -
\a_2-\a_3) -\cos (\a_1 +\a_2 +\a_3) + \cos (\a_1 -\a_2 -
\a_3-2\th_5) \right. \nonu \\
& +& \left. 
4\cos (\a_2 -\a_3 -\th_5) \cos \th_5 \sin \a_1 \right]
\tan \th_4.    
\nonu 
\eea
One checks the identities 
\bea
\hat{F}_k^{\,i} \, \hat{F}_j^{\,k} =-\delta_j^i, \qquad
\hat{F}^{[ij} \, \hat{F}^{kl} \, \hat{F}^{mn]} = \frac{1}{15} \, \eps6^{ijklmn}.
\nonu
\eea
The raising and lowering the indices are done by the 6-dimensional
metric on ${\bf S}^6$ (\ref{Met6d}).

The $\hat{T}_{ijk}$ tensor appearing in the 4-forms 
$F_{4mnp}$ and $F_{5mnp}$ is given by 
\bea
\hat{T}_{123} & = & \frac{1}{4} \cos \a_1 \sin^2 \a_1 \sin^3 \th_4
\sin (\a_2-\a_3 -\th_5) \sin^3 \th_6, \nonu \\
\hat{T}_{124} & = & \frac{1}{8} \sin^2 \th_4 \sin^3 \th_6 \left[
2 \cos \a_1 \cos \th_4 \cos (\a_2 -\a_3 -\th_5) + \sin \th_4 ( \cos
\th_5 \sin 2\a_1 \right. \nonu \\
& + & \left. \sin \th_5 ( \sin 2\a_2 -\cos 2\a_1 \sin 2(\al_3 +\th_5)))
\right],  
\nonu \\
\hat{T}_{125} & = &  \frac{1}{8} \sin^2 \th_4 \sin^3 \th_6 \left[
4 \cos \a_1 \sin^2 \a_1 \sin \th_4 \sin (\a_2 -\a_3 -\th_5) + \cos \th_4 ( 
\sin 2\a_1 \sin \th_5 \right. \nonu \\
& + & \left. \cos \th_5 ( -\sin 2\a_2 +\cos 2\a_1 \sin 2(\al_3 +\th_5)))
\right],  
\nonu \\
\hat{T}_{126} & = & \frac{1}{16} \sin^2 \th_4 \sin^2 \th_6 \left[
\cos 3\a_1  \cos (\a_2 -\a_3 -\th_5) \cos \th_6 \sin \th_4 \right.
\nonu \\
& + & \cos \a_1 \cos \th_6 ( - 4 \cos \th_4 \cos \th_5 \sin \a_1 +(5-2
\cos 2\a_1) \cos (\a_2 -\a_3-\th_5) \sin \th_4 ) \nonu \\
& - & 2 \cos \th_4 \cos \th_6 \sin \th_5 ( \sin 2\a_2 -\cos 2\a_1 \sin
2(\a_3 +\th_5)) \nonu \\
& + & \left. 2 (\cos 2\a_2 -\cos 2\a_1 \cos 2(\a_3 +\th_5)) \sin \th_6
\right], \nonu \\
\hat{T}_{134} & = & \frac{1}{8} \sin^2 \th_4 \sin^3 \th_6 \left[
2 \cos \a_1 ( \cos \th_4 \cos(\a_2 -\a_3 -\th_5)-\cos \th_5 \sin \a_1
\sin \th_4 ) \right.  
\nonu \\
& + & \left. \cos^2 \a_1 \sin 2\a_2 \sin \th_4 \sin \th_5 -
\sin \th_4 \sin \th_5 ( \sin^2 \a_1 \sin 2\a_2 + \sin 2(\a_3 +\th_5))  
\right],
\nonu \\ 
\hat{T}_{135} & = & \frac{1}{16} \cos \th_4 \sin^2 \th_4 \sin^3 \th_6
\left[ -\cos(2\a_1 -\th_5) + \cos (2\a_1 +\th_5) + \cos \th_5 (\sin
  2(\a_1-\a_2)\right.
  \nonu \\
&-& \left. \sin 2(\a_1+\a_2) + 2 \sin 2(\a_3 +\th_5)) \right], 
\nonu \\
\hat{T}_{136} & = & \frac{1}{16} \sin^2 \th_4 \sin^2 \th_6 \left[
4 \cos \a_1 \cos (\a_2 -\a_3-\th_5) \cos \th_6 \sin \th_4 \right.
 \nonu \\
& + & \cos \th_4 \cos \th_6 ( 2 \cos \th_5 \sin 2\a_1 + \sin \th_5 (
\sin 2(\a_1 -\a_2) -\sin 2(\a_1 +\a_2) \nonu \\
& + & \left. 2 \sin 2(\a_3 +\th_5))) + (\cos 2(\a_1-\a_2) + \cos 2(\a_1
+\a_2)- 2 \cos 2(\a_3 +\th_5)) \sin \th_6 \right],
\nonu \\ 
\hat{T}_{145} & = & \frac{1}{32} \sin \th_4 \sin^3 \th_6 \left[
2\cos (2\a_2 -\th_5) + \cos (2\a_2 -2\th_4 -\th_5) + \cos (2\a_2
+2\th_4-\th_5) \right. \nonu \\
& + & 2 \cos (2\a_2 +\th_5) + \cos (2\a_2 -2\th_4 +\th_5) - 2 \cos
(2\a_3 -2\th_4 +\th_5) + \cos (2\a_2 +2\th_4 +\th_5) 
\nonu \\
&-& 2 (\cos (2\a_3 + 2\th_4 +\th_5) + 2 \cos (2\a_3  +3 \th_5) +
2 ( \cos \a_1 \cos (\a_2 -\a_3 -\th_5) \nonu \\
& - &  \left. \cos (\a_2 +\a_3 +\th_5) \sin
\a_1 ) \sin 2\th_4 + 2 \sin^2 \th_4 ( - \cos \th_5 \sin 2\a_1 + \cos
2\a_1 \sin 2\a_2 \sin \th_5 ))) \right], 
\nonu \\ 
\hat{T}_{146} & = & \frac{1}{4} \sin \th_4 \sin^2 \th_6 \left[
( \cos 2\a_2 -\cos 2(\a_3 +\th_5)) \cos \th_6 \sin \th_5 \right. \nonu \\
& + & ( \cos \th_4 ( \sin 2\a_2 -\sin2(\a_3 +\th_5)) +
2\sin \th_4 ( - \cos \a_1 \cos (\a_2 -\a_3 -\th_5) \sin \th_5   
\nonu \\
& + & \left. \cos \th_5 \sin \a_1 \sin (\a_2 +\a_3 +\th_5))) \sin
  \th_6 \right],
\nonu \\
\hat{T}_{156} & = & \frac{1}{16} \sin \th_4 \sin^2 \th_6 \left[
\cos \th_6  ( 8 \cos^2 \th_4 \cos (\a_2 +\a_3 +\th_5) \sin \a_1
\right.
\nonu \\
& + & 8
\cos \a_1 \cos (\a_2 -\a_3 -\th_5) \sin^2 \th_4   
- \sin 2\th_4 ( \cos (2\a_2 -\th_5) + \cos (2\a_2 +\th_5) \nonu \\
&-& 2( \cos (2\a_3+\th_5) + \cos \th_5 \sin 2\a_1 -\cos 2\a_1 \sin
2\a_2 \sin \th_5 )))  \nonu \\
 & + & 2 (( \cos (2\a_1-2\a_2)+\cos (2\a_1 +2\a_2) -2 \cos(2\a_3
 +2\th_5)) \sin \th_4
\nonu \\ 
& + & \left. 4 \cos \th_4 ( \cos \a_1 \cos (\a_2-\a_3-\th_5) \cos \th_5 +
\sin \a_1 \sin \th_5 \sin (\a_2 +\a_3 +\th_5))) \sin 
\th_6 \right], 
\nonu \\
%
\hat{T}_{234} & = & \frac{1}{8} \sin 2\a_1 \sin^2 \th_4 \sin^3 \th_6 
\left[ - 2\cos \a_1 \cos \th_4 \sin (\a_2 -\a_3 -\th_5) \right. 
\nonu \\
& + &  \left. ( \cos 2\a_2 + \cos(2\a_3 +2\th_5) ) \sin \th_4 \sin \th_5 
\right], \nonu \\  
\hat{T}_{235} & = & -\frac{1}{8} \cos \th_4 \cos \th_5 
\sin 2\a_1 \sin^2 \th_4 \sin^3 \th_6 \left[
\cos 2\a_2 + \cos (2\a_3 +2\th_5)\right],
\nonu \\
\hat{T}_{236} & = & -\frac{1}{8} \sin 2\a_1 \sin^2 \th_4 \sin^2 \th_6 
\left[ 2 \cos\a_1 \cos \th_6 \sin \th_4 \sin (\a_2-\a_3-\th_5) \right.
 \nonu \\
& + & \left.
\cos \th_4 (\cos2\a_2 + \cos (2\a_3 +2\th_5)) \cos \th_6 \sin
\th_5 + (\sin 2\a_2 +\sin (2\a_3 + 2\th_5)) \sin \th_6 \right], 
\nonu \\ 
\hat{T}_{245} & = & -\frac{1}{4} \sin \th_4 \sin^3 \th_6 \left[ 
-2 \cos^2 \a_1 \sin \a_1 \sin 2\th_4 \sin (\a_2 -\a_3-\th_5) + ( 2
\cos^2 \th_4 \sin^2 \a_1 \right.  \nonu \\
&+ & ( \cos 2\a_2 + \cos (2\a_3 +2\th_5)) \sin 2\a_1 \sin^2 \th_4 )
\sin \th_5 \nonu \\
&+ & \left. \cos \a_1 ( 2 \cos^2 \th_4 \cos \th_5 \sin \a_1 \sin (2\a_3
+2\th_5) - \sin 2\th_4 \sin (\a_2 +\a_3 +\th_5) ) \right],
\nonu \\
\hat{T}_{246} & = & \frac{1}{8} \sin \th_4 \sin^2 \th_6 \left[
\cos \th_6 \sin 2\a_1 \sin 2\a_3 ( \sin \th_5 - \sin 3\th_5) \right.
\nonu \\
& + & 2 (
\cos \th_4 \cos (2\a_3 +2\th_5) \sin 2\a_1     
\nonu \\
& + & 2 \sin \a_1 \sin \th_4 \sin (\a_2 -\a_3 -\th_5) \sin \th_5 )
\sin \th_6 + 4\cos \th_5  ( \cos \th_6 ( \sin^2 \a_1 
\nonu \\
&-& \left. \cos 2\a_3 \sin 2\a_1 \sin^2 \th_5) - \cos \a_1 \cos 
(\a_2 +\a_3+\th_5) \sin \th_4 \sin \th_6)
\right], \nonu \\ 
\hat{T}_{256} & = & -\frac{1}{4} \sin^2 \th_6 \left[
4 \cos^2 \a_1 \cos \th_6 \sin \a_1 \sin^3 \th_4 \sin (\a_2
-\a_3-\th_5) \right. \nonu \\
&-& 2 \cos \a_1 \cos^2 \th_4 \cos \th_6 \sin \th_4 \sin (\a_2 + \a_3
+\th_5)
- \cos \th_4 \cos \th_6 \sin^2 \th_4 ( 2\sin \a_1 ( \nonu \\  
& -& \cos \a_1 \cos 2\a_2 + \sin \a_1 ) \sin \th_5  + \sin 2\a_1 \sin
(2\a_3 + \th_5) ) \nonu \\  
& + & \cos \a_1 \cos (\a_2 +\a_3+\th_5) \sin 2\th_4 \sin \th_5 \sin
\th_6  
 + ( \cos \th_5 \sin \a_1 \sin 2\th_4 \sin (\a_2 -\a_3 -\th_5) \nonu
 \\
& + & \left.
\sin 2\a_1 \sin^2 \th_4 ( \sin 2\a_2 + \sin (2\a_3 +2\th_5) )) \sin
\th_6 \right],
\nonu \\
\hat{T}_{345} & = & \frac{1}{4} \sin \th_4 \sin^3 \th_6 \left[
2\cos^2 \th_4 \sin \a_1 ( \cos \a_1 \cos \th_5 \sin 2\a_2 + \sin \a_1
\sin \th_5 ) \right. \nonu \\
& + & \left.
\cos \a_1 \sin 2\th_4 \sin (\a_2 +\a_3 +\th_5) \right], 
\nonu \\
\hat{T}_{346} & = & \frac{1}{4} \sin \th_4 \sin^2 \th_6 \left[
\cos \th_6 \sin 2\a_1 \sin 2\a_2 \sin \th_5 -\cos^2 \th_5 ( 2\cos \a_1
\cos (\a_2 +\a_3) \right. \nonu \\
& + & \cos (\a_2 -\a_3) \sin \a_1 ) \sin \th_4 \sin \th_6 + ( \cos
\th_4 \sin 2\a_1 \sin^2 \a_2 \nonu \\
& + & \sin \a_1 \sin \th_4 ( \cos (\a_2-\a_3)
(1+ \sin^2 \th_5) 
 +  \cos \a_2 \sin \a_3 \sin 2\th_5) \nonu \\
& + &
\cos \a_1 ( -2 \cos^2 \a_2 \cos \th_4 \sin \a_1 
 +  \sin (\a_2 +\a_3) \sin \th_4 \sin 2\th_5 )) \sin \th_6
\nonu \\
& - & \left.
2\cos \th_5 \sin \a_1 ( \cos \th_6 \sin \a_1 + \cos \a_3 \sin \a_2 \sin
\th_4  \sin \th_5 \sin \th_6)
\right], 
\nonu \\  
\hat{T}_{356} & = & \frac{1}{4} \sin^2 \th_6 \left[
\cos \th_4 \cos \th_6 ( -\sin^2 \th_4 ( \cos \th_5 \sin 2\a_1 \sin
2\a_2 + 2\sin^2 \a_1 \sin \th_5 ) \right. \nonu \\
& + & \cos \a_1 \sin 2\th_4 \sin (\a_2+\a_3+\th_5) ) + \sin 2\th_4 (
\cos \th_5 \sin \a_1 \sin (\a_2-\a_3-\th_5) \nonu \\
& - & \left. \cos \a_1 \cos (\a_2 +\a_3 +\th_5) \sin \th_5 ) 
\sin \th_6 \right],  
\nonu \\
%
\hat{T}_{456} & = & \frac{1}{8} \sin^2 \th_6 \left[
- 4 \cos^2 \a_1 \cos \th_5 \cos \th_6 \sin \th_4 -
4 \cos \th_6 \sin 2\a_1 \sin 2\a_2 \sin \th_4 \sin \th_5 \right. \nonu \\
& + & 4 \cos (\a_2 -\a_3) \cos^2 \th_5 \sin \a_1 \sin \th_6 \nonu \\
& + &
2 \sin \a_1 (\cos (\a_2-\a_3)(-1+ 2 \cos 2\th_4 + \cos 2\th_5) \nonu
\\
& - & 2 
\cos \a_2 \sin \a_3 \sin 2\th_5 ) \sin \th_6 - \cos \a_1 ( 8 \cos
\th_4
\cos (\a_2 +\a_3 +\th_5) \cos \th_6   \nonu \\
& + & ( -\cos (\a_1 +2\a_2 -2\th_4) + 2 \cos (\a_2 +\a_3 -2\th_4) +
\cos (\a_1 -2\a_2 +2\th_4) \nonu \\
& + & 2 \cos (\a_2 +\a_3 + 2\th_4) 
 -  \cos (\a_1 -2\a_2 -2\th_4) + \cos (\a_1 + 2\a_2 + 2\th_4) \nonu \\
& - &  4 \cos
 (\a_2 +\a_3 + 2\th_5) ) \sin \th_6) 
 +  4 \cos \th_5 (\cos \th_6 (1+\sin^2 \a_1) \sin \th_4 \nonu \\
& + & \left. 2\cos \a_3 
\sin \a_1 \sin \a_2 \sin \th_5 \sin \th_6 ) \right]. 
\nonu 
\eea
One has also various identities which can be used for checking the
11-dimensional Einstein-Maxwell equations. 
\bea
\hat{T}_{ij}^{\,\,\,m} \, \hat{F}_{mk} + \hat{T}_{kj}^{\,\,m}
\, \hat{F}_{mi}  
& = & 0, \nonu \\
 \hat{T}_{ij}^{\,\,m} \, \hat{T}_{mk}^{\,\,\,\,l} + 
\hat{T}_{kj}^{\,\,m} \, \hat{T}_{mi}^{\,\,\,\,l}  & = &
- \hat{F}_{ij} \, \hat{F}_k^{\,\,l} - \hat{F}_{kj} \, \hat{F}_i^{\,\,l} -  
\delta_i^l \, \g6_{jk} -\delta_k^l \, \g6_{ij} + 2 \delta_j^l \,
\g6_{ik}, 
\nonu \\
\nab6_k \, \hat{F}_{ij} & = & \hat{T}_{ijk}, \nonu \\
\nab6_k \, \hat{T}_{ij}^{\,\,l} & = & - \g6_{ki} \, \hat{F}_j^{\,\,l} +
\g6_{kj} \, \hat{F}_i^{\,\,l} - \delta_k^{l} \, \hat{F}_{ij}, 
\nonu \\
\nab6_k \, \hat{T}_{ij}^{\,\,k} & = & - 4 \, \hat{F}_{ij}, 
\nonu \\
\hat{T}_{imn} \, \hat{T}^{jmn} & = & 4 \, \delta_i^{j}, \nonu \\
\hat{T}_{ijk} \, \hat{F}^{jk} & = & 0, \nonu \\
\eps6_{ijkmnp} \, \hat{T}^{mnp} \, \hat{F}^{jk} & = & 0,
\nonu \\
\eps6^{ijkmnp} \, \nab6_i \hat{T}_{mnp} & = & 0, \nonu \\
\nab6_{[m} \, \hat{T}_{npq]} & = & 0. 
\nonu
\eea 
All these identities can be checked from the scalar product and the
vector cross product of two vectors in Cayley space ${\bf I}^7$.
According to the third equation, 
by taking the 6-dimensional covariant derivative on the almost complex
structure $\hat{F}_{ij}$, one obtains the $\hat{T}_{ijk}$ tensor.
In other words, the above expression for 
$\hat{T}_{ijk}$ can be determined via the
explicit form for $\hat{F}_{ij}$. We present them here for convenience.
In general, the second fundamental tensor of the hypersurface is
different from the metric tensor but for ${\bf S}^6$, they are
equivalent to each other.  We use this property all the times.

The $\hat{S}_{ijk}$ tensor appearing in the 4-forms 
$F_{4mnp}$ and $F_{5mnp}$ is given by 
\bea
\hat{S}_{123} & = & \frac{1}{64} \sin 2\a_1 \sin^3 \th_4 \sin^3 \th_6
\left[ -8 \cos \a_1 \cos (\a_2 +\a_3 +\th_5) \cos \th_6 
\right. \nonu \\
& - & \cos \th_4
( \cos (\a_1-\a_2-\a_3) + \cos (\a_1+\a_2+\a_3)+ 
2\cos \a_1 ( \cos (\a_2 +\a_3) \nonu \\ 
& - & \left. 2 \cos (\a_2 +\a_3+2\th_5)) -8
\cos (\a_2-\a_3-\th_5) \cos \th_5 \sin \a_1 ) 
\sin \th_6 \right], 
\nonu \\
\hat{S}_{124} & = & \frac{1}{8} \sin^2 \th_4 \sin^3 \th_6 \left[
\cos \th_6 ( \sin 2\a_1 \sin \th_4 \sin \th_5 - 2\cos \th_4 \sin \a_1
\sin ( \a_2 +\a_3 + \th_5) \right. \nonu \\
&+& 2 \cos (\a_2 +\a_3 +\th_5) \sin \a_1 \sin \th_5 \sin \th_6
+ \cos \th_5 ( \cos \th_6 \sin \th_4 ( \sin 2\a_2  \nonu \\ 
& + &  \left. \cos 2\a_1 \sin (2\a_3 +2\th_5) ) 
-2 \cos \a_1 \sin (
\a_2 -\a_3 -\th_5) \sin \th_6 ) \right],   
\nonu \\
\hat{S}_{125} & = & \frac{1}{16} \sin^2 \th_4 \sin^3 \th_6 \left[
2\cos \th_6 ( -2 \cos \a_1 \cos \th_4 \cos \th_5 \sin \a_1 \right.
\nonu \\
& - & 4 \cos^2
\a_1 \cos (\a_2 +\a_3 + \th_5) \sin \a_1 \sin \th_4  + \cos \th_4
\sin \th_5 ( \sin 2\a_2 + \cos 2\a_1 \sin (2\a_3 +2\th_5))) \nonu 
\\ 
&+ & (-2 \cos 2\a_2 \cos^2 \th_4 -2 \cos 2\a_1 \cos \th_4 (
\cos \th_4 \cos (2\a_3+2\th_5) \nonu \\
& + & ( \cos \a_1 \cos (\a_2-\a_3-2\th_5)
\nonu \\ 
&- & \cos (\a_2 +\a_3+2\th_5) \sin \a_1 ) \sin \th_4 )
+ ( \cos \a_2 \cos \a_3 ( \cos \a_1 -\sin \a_1)^3 \nonu \\
& + & \left.
(\cos \a_1 + \sin \a_1)^3 \sin \a_2 \sin \a_3 ) \sin 2\th_4 )
\sin \th_6 \right],  
\nonu \\
\hat{S}_{126} & = & -\frac{1}{8} \sin 2\a_1 \sin^3 \th_4  \sin^2 \th_6
\left[ \cot \th_4 ( \cos \a_2 \cos \th_5 \csc \a_1 \sec \a_1 \sin \a_2
  \right. \nonu \\
& + & \left.
\sin \th_5 + \cos \th_5 \cot 2\a_1 \sin (2\a_3 +2\th_5)) +
\sec \a_1 \sin (\a_2 +\a_3 +\th_5) \right],   
\nonu \\
\hat{S}_{134} & = & \frac{1}{24} \sin^2 \th_4 \sin^3 \th_6 
\left[ 3 \cos (2\a_3 +2\th_5) \cos \th_6 \sin \th_4 \sin \th_5 + \cos \th_4 \cos \th_6 
( \csc \a_1 \right. \nonu \\
& + & 4 \sin \a_1) \sin (\a_2 +\a_3 +\th_5)
-  ( \cos \th_5 ( 4\cos \a_1 + \sec \a_1 ) \sin (\a_2 -\a_3 -\th_5)
\nonu \\
& + & \cos (\a_2 +\a_3 +\th_5) ( \csc \a_1 + 4 \sin \a_1 ) \sin \th_5 )
\sin \th_6 
 +  3 \cos \th_6 \sin \th_4 ( \sin (2\a_3 +\th_5) 
\nonu \\
& + & 
\sin \th_5 \tan \a_1 ) + \cos 2\a_1 ( -\cos \th_4 \cos \th_6 \csc \a_1
\sin (\a_2 +\a_3 +\th_5) \nonu \\
& + & \cos \th_5 (3 \cos \th_6 \sin 2\a_2 \sin \th_4 -\sec \a_1 
\sin (\a_2-\a_3-\th_5) \sin \th_6 ) \nonu \\
& + & \left. \sin \th_5 (\cos (\a_2 +\a_3 + \th_5) \csc \a_1 \sin \th_6 +
3 \cos \th_6 \sin \th_4 \tan \a_1) ) \right],  
\nonu \\
\hat{S}_{135} & = & -\frac{1}{32} \sin^3 \th_6 \left[
( \cos 2\a_1 \cos 2\a_2 + \cos (2\a_3 +2\th_5) ) \sin^2 2\th_4 
\sin \th_6  \right. \nonu \\
&- & 4 \cos \th_4 \sin^2 \th_4 ( \cos \th_5 ( -\cos \th_6 \sin 2\a_1 +
2 \cos (\a_2 +\a_3 +\th_5) \sin \a_1 \sin \th_4 \sin \th_6 ) \nonu \\
&+& 
\sin \th_5 ( \cos \th_6 ( \cos 2\a_1 \sin 2\a_2 + \sin (2\a_3 + 
2\th_5) ) \nonu \\
& - & \left.
2 \cos \a_1 \sin \th_4 \sin (\a_2 -\a_3 -\th_5) 
\sin \th_6 ) ) \right],  
\nonu \\
\hat{S}_{136} & = & \frac{1}{16} \sin^2 \th_4 \sin^2 \th_6 
\left[ -\cos \th_4 ( \cos (2\a_1 -\th_5) -\cos (2\a_1 +\th_5) +
2\cos \th_5 ( \cos 2\a_1 \sin 2\a_2 \right. \nonu \\
&  + & \left. \sin (2\a_3 + 2\th_5) ) )
 +  4 \sin \a_1 \sin \th_4 \sin (\a_2 +\a_3 +\th_5) 
\right], \nonu \\
%
\hat{S}_{145} & = & -\frac{1}{48} \sin \th_4 \sin^3 \th_6 \left[
12 \cos 2\a_2 \cos^2 \th_4 \cos \th_6 \sin \th_5 + 8 \cos (2\a_3 +
2\th_5) \cos \th_6 \sin \th_5 \right.
\nonu \\
& - & 4 \cos (2\a_3 +2\th_5) \cos \th_6 \cot^2 2\a_1 \sin \th_5 
+ 4 \cos (2\a_3 +2\th_5) \cos \th_6 \csc^2 2\a_1 \sin \th_5 \nonu \\
& + & 2 \cos 2\a_3 \cos \th_6 \sin^2 \th_4 \sin \th_5 + 8 \cos \th_6 
\sin 2\a_1 \sin^2 \th_4 \sin \th_5 \nonu \\
& + & \cos \th_6 \csc \a_1 \sin 2\th_4 \sin (\a_2 +\a_3 +\th_5)
+ 10 \cos \th_6 \sin \a_1 \sin 2\th_4 \sin (\a_2 +\a_3 + \th_5) \nonu
\\
& + & 10 \cos \th_6 \sin^2 \th_4 \sin (2\a_3 +\th_5) + 6 \cos \th_4 
\sin 2\a_2 \sin \th_6 + 2 \sec \th_4 \sin 2\a_2 \sin \th_6
\nonu \\
& - & 4 \cot^2 2\a_1 \sec \th_4 \sin 2\a_2 \sin \th_6 + 4 \csc^2 2\a_1
\sec \th_4 \sin 2\a_2 \sin \th_6 \nonu \\
& - & 4 \cos (\a_2 +\a_3 +\th_5) \csc \a_1 \sin \th_4 \sin \th_5 \sin
\th_6
-16 \cos (\a_2 +\a_3 +\th_5) \sin \a_1 \sin \th_4 \sin \th_5 \sin \th_6
\nonu \\
& + & 6 \cos \th_4 \sin (2\a_3 +2\th_5) \sin \th_6 + 2 \sec \th_4 
\sin (2\a_3 +2\th_5) \sin \th_6 \nonu \\
& - & 4 \cot^2 2\a_1 \sec \th_4 \sin (2\a_3 +2\th_5) \sin \th_6 + 
4 \csc^2 2\a_1 \sec \th_4 \sin (2\a_3 +2\th_5) \sin \th_6 \nonu \\
&+ & 2 \cos \th_5 \sin \th_4 ( \cos \th_6 \sin 2\a_3 \sin \th_4 -
3 ( 2\cos \a_1 + \sec \a_1 ) \sin (\a_2 -\a_3-\th_5) \sin \th_6 ) \nonu
\\
&+ & 4 \cos \th_6 \sin^2 \th_4 \sin \th_5 \tan \a_1 + \cos 2\a_1 (
4 \cos (\a_2 +\a_3 +\th_5) \csc \a_1 \sin \th_4 \sin \th_5 \sin 
\th_6   \nonu \\
& + &  6 \cos \th_5 \sin \th_4 ( 2 \cos \th_6 \sin 2\a_2 \sin \th_4 -
\sec \a_1 \sin (\a_2 -\a_3 -\th_5) \sin \th_6 ) \nonu \\
& + &  \cos \th_6 ( -\csc \a_1 \sin 2\th_4 \sin (\a_2 +\a_3 +\th_5) +
4 \sin^2 \th_4 \sin \th_5 \tan \a_1 ) ) \nonu \\
& + & \cos \th_6 \csc^2 \a_1 \sec \a_1 \sin (\a_2-\a_3 -\th_5) 
\tan \th_4 - 6 \sin 2\a_2 \sin \th_4 \sin \th_6 \tan \th_4 \nonu \\
& - & 6 \sin \th_4 \sin (2\a_3 + 2\th_5) \sin \th_6 \tan \th_4 \nonu
\\
& + & \left. 2 \cos \a_1 \cos \th_6 \sin (\a_2 -\al_3-\th_5) (
5 \sin 2\th_4 - 2 ( \cot^2 2\a_1 + \sin^2 \th_4 ) 
\tan \th_4  ) \right],   \nonu \\
\hat{S}_{146} & = & \frac{1}{4} \cos \th_5 \sin \th_4 \sin^2 \th_6 
\left[ \cos 2\a_2 + \cos (2\a_3 + 2\th_5) 
\right], \nonu \\
\hat{S}_{156} & = & \frac{1}{8}  \sin \th_4 \sin^2 \th_6 \left[
-4 \cos^2 \a_1 \cos \a_2 \cos \th_4 \cos \th_5 \sin \a_2 \sin 
\th_4 \right. \nonu \\
& + & \sin 2\th_4 ( \cos \th_5 ( \sin^2 \a_1 \sin 2\a_2 - 2 
\cos \a_3 \sin \a_3 ) + (\cos 2\a_2 -\cos 2\a_3 ) \sin \th_5  )
\nonu \\ 
&- &  4 \cos \a_1 \cos \th_4 ( \cos \th_4 \sin 
(\a_2 -\a_3 -\th_5) + \sin \a_1 \sin \th_4 \sin \th_5 )
\nonu \\
& + & \left. 
4 \sin \a_1 \sin^2 \th_4 \sin (\al_2 + \a_3 +\th_5) \right],  
\nonu \\
\hat{S}_{234} & = & \frac{1}{8}  \sin 2\a_1 \sin^2 \th_4 \sin^3 \th_6 
\left[ -2 \cos \th_4 \cos (\a_2 +\a_3 +\th_5) \cos \th_6 \sin \a_1
  \right. \nonu \\
& +  & \cos \th_5 (\cos 2\a_2 - \cos (2\a_3 + 2\th_5) ) \cos \th_6 
\sin \th_4 - 2 ( \cos \a_1 \cos (\a_2 -\a_3 -\th_5) \cos \th_5 \nonu
\\
& + & \left. \sin \a_1 \sin \th_5 \sin (\a_2 +\a_3 +\th_5) )
\sin \th_6 \right],  
\nonu \\
\hat{S}_{235} & = & \frac{1}{8} \cos \th_4 \sin 2\a_1 \sin^2 \th_4
\sin^3
\th_6 \left[ (\cos 2\a_2 -\cos(2\a_3 +2\th_5 )) \cos \th_6 \sin
  \th_5 \right. \nonu \\
&+ & ( \cos \th_4 (\sin 2\a_2 -\sin (2\a_3 +2\th_5) )  
+ 2 \sin \th_4 ( -\cos \a_1 \cos (\a_2-\a_3 -\th_5) \sin \th_5 \nonu \\
&+ & \left. \cos \th_5 \sin \a_1 \sin (\a_2 +\a_3 +\th_5) ) ) 
\sin \th_6 \right],   
\nonu \\
\hat{S}_{236} & = & -\frac{1}{8} \sin 2\a_1 
\sin^2 \th_4  \sin^2 \th_6 \left[
\cos \th_4 \cos \th_5 ( \cos 2\a_2 -\cos (2\a_3 +2\th_5)) \right.
\nonu \\
& + & \left. 2\cos (\a_2 +\a_3 +\th_5) \sin \a_1 \sin \th_4   
\right], \nonu \\
%
\hat{S}_{245} & = & \frac{1}{8} \sin \th_4 \sin^3 \th_6 \left[
-\cos \th_6 ( 2 \cos (\a_2 -\a_3 -\th_5) \sin \a_1 \sin 2\th_4 \right.
\nonu \\
& + & \sin 2\a_1 ( 2 \cos 2\a_2 \cos \th_5 \sin^2 \th_4 + \sin 2\a_3 
\sin 3\th_5 ) ) \nonu \\
& + & \cos 2\a_3 \sin 2\a_1 ( ( -\cos 2\th_4 
\cos \th_5 
 +   \cos 3\th_5 ) \cos \th_6 + 2\cos \th_4 \cos 2\th_5 \sin \th_6 )
\nonu \\
& - &  4 \cos^2 \a_1 \cos \th_5 ( \cos^2 \th_4 \cos \th_6 -
2 \sin \a_1 \sin (\a_2-\a_3) \sin \th_4 \sin \th_5 \sin \th_6 )
\nonu \\
& + & 2 \cos \a_1 ( 2 \cos (\a_2 +\a_3 +\th_5) \cos \th_6 \sin^2 
\a_1  \sin 2\th_4 \nonu \\
& +& 2 \cos (\a_2 -\a_3) \cos^2 \th_5 \sin 2\a_1 \sin \th_4 \sin \th_6 
+ \sin \a_1 \sin \th_5 ( \cos 2\th_4 \cos \th_6 \sin 2\a_3 \nonu \\
& + & \left. 4 ( -\cos \th_4 \cos \th_5 \sin 2\a_3 + \sin \a_1 \sin \th_4 
\sin (\a_2 +\a_3 +\th_5) ) \sin \th_6 ) ) 
\right],   \nonu \\
\hat{S}_{246} & = & \frac{1}{2} \cos \a_1 \sin \th_4 \sin^2 \th_6 
\left[ -\cos \a_1 \sin \th_5 + \cos \th_5 \sin \a_1 \sin (2\a_3
  +2\th_5)
\right],  \nonu \\
\hat{S}_{256} & = & \frac{1}{8} \sin 2\a_1 \sin \th_4 
\sin^2 \th_6  \left[
-2 \cos^2 \th_4 \cos (\a_2 -\a_3 -\th_5) \sec \a_1 \right. \nonu \\
& - &  4 \cos (\a_2
+\a_3 +\th_5) \sin \a_1 \sin^2 \th_4 \nonu \\
& + & \left. 
( \cos (2\a_3 +\th_5) + \cos \th_5 (-\cos 2\a_2 + \cot \a_1 ))
\sin 2\th_4 \right],
\nonu \\
\hat{S}_{345} & = & \frac{1}{4} \cos \th_4 \sin 2\a_1 \sin \th_4
\sin^3 \th_6  \left[
\cos (\a_2 -\a_3 -\th_5) \cos \th_6 \sec \a_1 \sin \th_4 \right.
 \nonu \\
&-& \left.
\cos \th_4 \cos \th_6 ( \cos \th_5 \cot \a_1 + \sin 2\a_2 \sin 
\th_5 ) + \cos 2\a_2 \sin \th_6 
\right], \nonu \\
\hat{S}_{346} & = & \frac{1}{4} \sin 2\a_1 \sin \th_4 \sin^2 \th_6 
\left[ \cos \th_5 \sin 2\a_2 -\cot \a_1 \sin \th_5 \right],
\nonu \\
\hat{S}_{356} & = & \frac{1}{4} \cos \th_4 \sin 2\a_1 \sin^2 \th_4 
\sin^2 \th_6 \left[
\cos \th_5 \cot \a_1 \right. \nonu \\
& + & \left. \cos (\a_2 -\a_3-\th_5) \cot \th_4 \sec \a_1 
+ \sin 2\a_2 \sin \th_5  \right],  
\nonu \\
\hat{S}_{456} & = & \frac{1}{2} \sin^2 \th_6 \left[
2 \cos \a_2 \cos \th_4 \cos \th_5 \sin \a_1 \sin \a_3 \right. \nonu \\
& + & \sin \th_4 ( -\cos \th_5 \sin 2\a_1 \sin 2\a_2 + 2\cos^2 \a_1
\sin \th_5 ) \nonu \\
& + & \left. 2 \cos \th_4 \sin \a_1 ( -\cos \a_3 \sin (\a_2-\th_5) 
+ \sin \a_2 \sin \a_3 \sin \th_5 ) \right].
\nonu
\eea
Let us also present the other identities which contain the above
$\hat{S}_{ijk}$ 
tensor.
\bea
\hat{S}_{imn} \, \hat{S}^{jmn} & = & 4 \, \delta_i^{j},
\nonu \\
\hat{S}_{imn} \, \hat{T}^{jmn} & = 
& -\frac{1}{2} \, \g6_{ik} \, \eps6^{jkmnpq} \, \hat{F}_{mn} \, \hat{F}_{pq} = 
4 \hat{F}_i^j,
\nonu \\
\hat{S}_{ijk} \, \hat{F}^{jk} & = & 0, \nonu \\
\nab6_i \, \hat{S}^{ijk} & = & 0,
\nonu \\
\hat{T}_{imn} \, \hat{S}^{jmn} + \hat{S}_{imn} \, \hat{T}^{jmn}  & = & 0.
\nonu 
\eea
There is no $G_2$-invariant vector.
As mentioned, there exists a relation  (\ref{cp2condition}) 
which shows how the above tensor $\hat{S}_{ijk}$
can be obtained from the previous two tensors.


\section{The Ricci tensor  }

The Ricci tensor appearing in (\ref{rtilde}) is given by explicitly
\bea
\widetilde{R}_{1}^{\,\,1} & = & R_{1}^{\,\,1}, \qquad
\widetilde{R}_{2}^{\,\,2}  =  R_{2}^{\,\,2}\ (= \widetilde{R}_{1}^{\,\,1}), \qquad
\widetilde{R}_{3}^{\,\,3} =  R_{3}^{\,\,3} (= \widetilde{R}_{1}^{\,\,1}),  \nonu \\
\widetilde{R}_{4}^{\,\,4}  &  = & R_{4}^{\,\,4}, \nonu \\
\widetilde{R}_{4}^{\,\,5} & = & -\left[\frac{
\cos \mu \, \cos
  \psi}
{\sqrt{1-\cos^2 \psi \, \sin^2
    \mu} } \right] R_{4}^{\,\,5}, \nonu \\
\widetilde{R}_{4}^{\,\,10}  & = &  -\left[\frac{
\csc \mu \, \sin
  \psi}
{\sqrt{1-\cos^2 \psi \, \sin^2
    \mu} } \right] R_{4}^{\,\,5}, \nonu \\
\widetilde{R}_{4}^{\,\,11}  & = &  \left[\frac{ \csc \mu \, \sin
  \psi}
{\sqrt{1-\cos^2 \psi \, \sin^2
    \mu} } \right] R_{4}^{\,\,5} (= -
\widetilde{R}_{4}^{\,\,10}),
\nonu \\
\widetilde{R}_{5}^{\,\,4} & = & -\left[
\frac{\cos \mu \,\cos \psi}{\sqrt{1-\cos^2 \psi \, \sin^2
    \mu} } \right] R_{5}^{\,\,4},
 \nonu \\
\widetilde{R}_{5}^{\,\,5} & = & \left[\frac{\cos^2 \mu \, \cos^2 \psi }
{1-\cos^2 \psi \, \sin^2
    \mu } \right] R_{5}^{\,\,5} + \left[\frac{\sin^2 \psi}{1-\cos^2 \psi \, \sin^2
    \mu} \right] R_{11}^{\,\,11}, \nonu \\
\widetilde{R}_{5}^{\,\,10} & = & \left[\frac{\cot \mu \, \sin \psi \, \cos \psi }
{1-\cos^2 \psi \, \sin^2
    \mu } \right] R_{5}^{\,\,5} - \left[\frac{\cot \mu \, \sin \psi \, 
\cos \psi}{1-\cos^2 \psi \, \sin^2
    \mu} \right] R_{11}^{\,\,11},  
\nonu \\
\widetilde{R}_{5}^{\,\,11} & = & -\left[\frac{\cot \mu \, \sin \psi \, \cos \psi }
{1-\cos^2 \psi \, \sin^2
    \mu } \right] R_{5}^{\,\,5} + \left[\frac{\cot \mu \, \sin \psi \, 
\cos \psi}{1-\cos^2 \psi \, \sin^2
    \mu} \right] R_{11}^{\,\,11} (=  -\widetilde{R}_{5}^{\,\,10}),
\nonu \\
\widetilde{R}_{6}^{\,\,6} & = & R_{6}^{\,\,6}, \qquad 
\widetilde{R}_{7}^{\,\,7}  =  R_{7}^{\,\,7} (= \widetilde{R}_{6}^{\,\,6}), \qquad
\widetilde{R}_{8}^{\,\,8}=   R_{8}^{\,\,8} (=
\widetilde{R}_{6}^{\,\,6}), \nonu \\
\widetilde{R}_{9}^{\,\,9} & = &  R_{9}^{\,\,9} (= \widetilde{R}_{6}^{\,\,6}), \qquad
\widetilde{R}_{10}^{\,\,10} =  R_{10}^{\,\,10} (= \widetilde{R}_{6}^{\,\,6}), 
\nonu \\
\widetilde{R}_{11}^{\,\,4} & = &  \left[\frac{\sin \mu \, \sin \psi}
{\sqrt{1-\cos^2 \psi \, \sin^2
    \mu}} \right] R_{5}^{\,\,4}, \nonu \\
 \widetilde{R}_{11}^{\,\,5}  & = &
 -\left[\frac{\sin \mu\, \cos \mu \, \sin \psi \, \cos \psi }
{1-\cos^2 \psi \, \sin^2
    \mu } \right] R_{5}^{\,\,5} + \left[\frac{\sin \mu \, \cos \mu \, \sin \psi \, 
\cos \psi}{1-\cos^2 \psi \, \sin^2
    \mu} \right] R_{11}^{\,\,11},
\nonu \\
 \widetilde{R}_{11}^{\,\,10}  & = &
 -\left[\frac{\sin^2 \psi }
{1-\cos^2 \psi \, \sin^2
    \mu } \right] R_{5}^{\,\,5} + R_{10}^{\,\, 10} 
- \left[\frac{\cos^2 \mu \, \cos^2 \psi }{1-\cos^2 \psi \, \sin^2
    \mu} \right] R_{11}^{\,\,11},
\nonu \\
 \widetilde{R}_{11}^{\,\,11}  & = &
 \left[\frac{\sin^2 \psi }
{1-\cos^2 \psi \, \sin^2
    \mu } \right] R_{5}^{\,\,5} + \left[\frac{\cos^2 \mu \, 
\cos^2 \psi}{1-\cos^2 \psi \, \sin^2
    \mu} \right] R_{11}^{\,\,11}.
\nonu
\eea
Here the Ricci tensor $R_{M}^{\,\, N}$ in the right hand side 
above is for $G_2$-invariant case we presented in the Appendix A and 
the old variable $\theta$ 
should be replaced by the new variables $(\mu, \psi)$ 
through (\ref{rel}). Note that there are off-diagonal components
$(4,10), (4,11), (5,10), (5,11), (11,4), (11,5)$, and $(11,10)$ while
there exist only $(4,5)$ and $(5,4)$ components in $G_2$-invariant flow.
One also obtains these components from the 11-dimensional metric
(\ref{11frames}) 
directly.

\section{The 4-forms}

The 4-forms appearing in (\ref{ftilde}) are given by 
\bea
\widetilde{F}_{1234} & = & F_{1234}, \qquad
\widetilde{F}_{1235}   =  -\left[\frac{\cos \mu \, \cos \psi}
{\sqrt{1-\cos^2 \psi \, \sin^2
    \mu}} \right] F_{1235}, \nonu \\
\widetilde{F}_{123\,11}  & = &  \left[\frac{\sin \mu \, \sin \psi}
{\sqrt{1-\cos^2 \psi \, \sin^2
    \mu}} \right] F_{1235},
\nonu \\
\widetilde{F}_{45mn} & = & -\left[\frac{\cos \mu \, \cos \psi}
{\sqrt{1-\cos^2 \psi \, \sin^2
    \mu}} \right] F_{45mn} - \left[ \frac{\sin \psi}
{1-\cos^2 \psi \, \sin^2
    \mu} \right] F_{4mn\, 11}, (m, n=6, \cdots,  10),
\nonu \\
\widetilde{F}_{45m\,11} & = & -\left[\frac{\cos \mu \, \cos \psi}
{\sqrt{1-\cos^2 \psi \, \sin^2
    \mu}} \right] F_{45m\, 10} + \left[\frac{\sin \mu }
{\sqrt{1-\cos^2 \psi \, \sin^2 \mu}} \right] F_{45m\, 11} \nonu \\
& - & \left[\frac{\sin \psi}{1-\cos^2 \psi \, \sin^2
    \mu} \right] F_{4m\,10\,11}, \quad (m =6, \cdots, 9),
\nonu \\
\widetilde{F}_{45\,10\,11} & = & 
\left[\frac{\sin \mu}
{\sqrt{1-\cos^2 \psi \, \sin^2 \mu}} \right] F_{45\,10\,11},
\qquad
\widetilde{F}_{4mnp}  =  F_{4mnp}, (m, n, p =6, \cdots, 10),
\nonu \\
\widetilde{F}_{4mn\,11} & = & \left[\frac{\sin \mu \, \sin \psi}
{\sqrt{1-\cos^2 \psi \, \sin^2
    \mu}} \right] F_{45mn} + F_{4mn\, 10} - \left[\frac{\sin \mu\, \cos \mu \cos
  \psi}
{1-\cos^2 \psi \, \sin^2
    \mu} \right] F_{4mn\,11}, \nonu \\
 & & (m, n =6, \cdots, 9),
\nonu \\
\widetilde{F}_{4m\,10\,11} & = & \left[\frac{\sin \mu \, \sin \psi}
{\sqrt{1-\cos^2 \psi \, \sin^2
    \mu}} \right] F_{45m\,10}  - \left[\frac{\sin \mu \, \cos \mu\, \cos 
\psi}{1-\cos^2 \psi \, \sin^2
    \mu} \right] F_{4m\,10\,11}, (m =6, \cdots, 9),
\nonu \\
\widetilde{F}_{5mnp} & = &  -\left[\frac{\cos \mu \, \cos \psi}
{\sqrt{1-\cos^2 \psi \, \sin^2
    \mu}} \right] F_{5mnp} +   \left[\frac{\sin \psi}{1-\cos^2 \psi \, \sin^2
    \mu} \right] F_{mnp\,11}, (m, n, p =6, \cdots, 10),
\nonu \\
\widetilde{F}_{5mn\,11} & = & -\left[\frac{\cos \mu \, \cos \psi}
{\sqrt{1-\cos^2 \psi \, \sin^2
    \mu}} \right] F_{5mn\,10} + \left[\frac{\sin \mu }
{\sqrt{1-\cos^2 \psi \, \sin^2 \mu}} 
\right] F_{5mn\, 11} \nonu \\
&+ & \left[\frac{\sin
  \psi}
{1-\cos^2 \psi \, \sin^2
    \mu} \right] F_{mn\,10\,11}, \quad (m, n =6, \cdots, 9),
\nonu \\
\widetilde{F}_{5m\,10\,11} & = & \left[\frac{\sin \mu }
{\sqrt{1-\cos^2 \psi \, \sin^2 \mu}} 
\right] F_{5m\,10\,11}, \quad (m = 6, \cdots, 9),
\nonu \\
\widetilde{F}_{mnpq} & = &  F_{mnpq}, \quad (m, n, p, q =6, \cdots, 10),
\nonu \\
\widetilde{F}_{mnp\,11} & = & -\left[\frac{\sin \mu \, \sin \psi}
{\sqrt{1-\cos^2 \psi \, \sin^2
    \mu}} \right] F_{5mnp} +  F_{mnp\, 10} - \left[\frac{\sin \mu \, \cos
 \mu\, \cos  \psi}
{1-\cos^2 \psi \, \sin^2
    \mu} \right] F_{mnp\,11}, \nonu \\
& & (m, n, p =6, \cdots, 9),
\nonu \\
\widetilde{F}_{mn\,10\,11} & = & -\left[\frac{\sin \mu \, \sin \psi}
{\sqrt{1-\cos^2 \psi \, \sin^2
    \mu}} \right] F_{5mn\,10}  - \left[\frac{\sin \mu \, \cos
 \mu\, \cos  \psi}
{1-\cos^2 \psi \, \sin^2
    \mu} \right] F_{mn\,10\,11}, (m, n=6, \cdots, 9).
\nonu
\eea
Here the 4-forms $F_{MNPQ}$ in the right hand side 
is for $G_2$-invariant case and is given by (\ref{fst2}) together
with the Appendices $B$ and $D$. Note that 
the old variables $(\theta, \theta_5, \theta_6)$ 
should be replaced by the new variables $(\mu, \phi, \psi)$ 
through (\ref{rel}). 
The 4-form $\widetilde{F}_{123\,11}$ is new. 

The 4-forms with upper indices are given by
\bea
\widetilde{F}^{1234} & = & F^{1234}, \qquad
\widetilde{F}^{1235}   =  -\left[\frac{\cos \mu \, \cos \psi}
{\sqrt{1-\cos^2 \psi \, \sin^2
    \mu}} \right] F^{1235}, \nonu \\
\widetilde{F}^{123\,10}  & = &  -\left[\frac{\csc \mu \, \sin \psi}
{\sqrt{1-\cos^2 \psi \, \sin^2
    \mu}} \right] F^{1235}, \qquad
\widetilde{F}^{123\,11}   =   \left[\frac{\csc \mu \, \sin \psi}
{\sqrt{1-\cos^2 \psi \, \sin^2
    \mu}} \right] F^{1235},
\nonu \\
\widetilde{F}^{45mn} & = & -\left[\frac{\cos \mu \, \cos \psi}
{\sqrt{1-\cos^2 \psi \, \sin^2
    \mu}} \right] F^{45mn} - \left[\sin \psi
 \right] F^{4mn\, 11}, \qquad (m, n=6, \cdots,  9),
\nonu \\
\widetilde{F}^{45m\,10} & = & -\left[\frac{\cos \mu \, \cos \psi}
{\sqrt{1-\cos^2 \psi \, \sin^2
    \mu}} \right] F^{45m\, 10} - \left[\csc \mu \, \sqrt{1-\cos^2 \psi \, \sin^2
    \mu}  \right] F^{45m\, 11} \nonu \\
& - & \left[\sin \psi  \right] F^{4m\,10\,11}, \qquad (m =6, \cdots, 9),
\nonu \\
\widetilde{F}^{45m\,11} & = & \left[
\csc \mu \, \sqrt{1-\cos^2 \psi \, \sin^2
    \mu} \right] F^{45m\, 11}, \qquad (m =6, \cdots, 10),
\nonu \\
\widetilde{F}^{4mnp}   & = &  F^{4mnp}, \qquad (m, n, p =6, \cdots, 9),
\nonu \\
\widetilde{F}^{4mn\,10} & = & F^{4mn\, 10} + \left[
\cos \psi \, \cot \mu  \right] F^{4mn\, 11} 
 -  \left[\frac{\csc \mu \,  \sin \psi}{\sqrt{1-\cos^2 \psi \, \sin^2
    \mu}}  \right] F^{45mn}, \qquad (m, n =6, \cdots, 9),
\nonu \\
\widetilde{F}^{4mn\,11} & = & \left[\frac{\csc \mu \, \sin \psi}
{\sqrt{1-\cos^2 \psi \, \sin^2
    \mu}} \right] F^{45mn}  - \left[\cos
  \psi \, \cot \mu  \right] F^{4mn\,11}, \qquad (m, n =6, \cdots, 10),
\nonu \\
\widetilde{F}^{5mnp} & = &  -\left[\frac{\cos \mu \, \cos \psi}
{\sqrt{1-\cos^2 \psi \, \sin^2
    \mu}} \right] F^{5mnp} +   \left[\sin \psi \right] 
F^{mnp\,11}, \qquad (m, n, p =6, \cdots, 9),
\nonu \\
\widetilde{F}^{5mn\,10} & = & -\left[\frac{\cos \mu \, \cos \psi}
{\sqrt{1-\cos^2 \psi \, \sin^2
    \mu}} \right] F^{5mn\, 10} - \left[\csc \mu \, \sqrt{1-\cos^2 \psi \, \sin^2
    \mu}  \right] F^{5mn\, 11} \nonu \\
& + & \left[\sin \psi  \right] F^{mn\,10\,11}, \qquad (m, n =6, \cdots, 9),
\nonu \\
\widetilde{F}^{5mn\,11} & = & \left[\csc \mu \, 
\sqrt{1-\cos^2 \psi \, \sin^2 \mu} 
\right] F^{5mn\, 11}, \qquad (m, n =6, \cdots, 10),
\qquad
\widetilde{F}^{6789}  =   F^{6789},
\nonu \\
\widetilde{F}^{mnp\,10} & = & F^{mnp\, 10} + \left[
\cos \psi \, \cot \mu  \right] F^{mnp\, 11} 
 +  \left[\frac{\csc \mu \,  \sin \psi}{\sqrt{1-\cos^2 \psi \, \sin^2
    \mu}}  \right] F^{5mnp}, 
\qquad (m, n, p =6, \cdots, 9),
\nonu \\
\widetilde{F}^{mnp\,11} & = & -\left[\frac{\csc \mu \, \sin \psi}
{\sqrt{1-\cos^2 \psi \, \sin^2
    \mu}} \right] F^{5mnp}  - \left[\cos  \psi \, \cot \mu
 \right] F^{mnp\,11}, \qquad (m, n, p =6, \cdots, 10).
\nonu
\eea
The 4-forms $\widetilde{F}^{123\,10}$ and $\widetilde{F}^{123\,11}$
are 
new, compared to the $G_2$-invariant flow.

\section{The left hand side of Maxwell equation}

The Maxwell equations can be summarized as follows.
Let us introduce the notation 
\bea
\frac{1}{2} \, E \,  
\widetilde{\nabla}_M  \, \widetilde{F}^{MNPQ} \equiv (NPQ),
\nonu
\eea
where we ignore the tilde in $(NPQ)$ for simplicity and 
present all the nonzero components of left hand side of Maxwell
equations in terms of the 4-forms in $G_2$-invariant case
\bea
(123) & = & -F_{49\,10\,11} F_{5678} + F_{48\,10\,11} F_{5679}
-F_{489\,11} F_{567\,10} + F_{489\,10} F_{567\,11}- F_{47\,10\,11} F_{5689}
 \nonu \\
&+ & F_{479\,11} F_{568\,10}- F_{479\,10} F_{568\,11} - F_{478\,11}
F_{569\,10} + 
F_{478\,10} F_{569\,11} - F_{4789} F_{56\,10\,11} \nonu \\  
&+ & F_{46\,10\,11} F_{5789} - F_{469\,11} F_{578\,10} + F_{469\,10}
F_{578\,11}  
+ F_{468\,11} F_{579\,10} - F_{468\,10} F_{579\,11} \nonu \\
&+ & F_{4689} F_{57\,10\,11} - F_{467\,11} F_{589\,10} + F_{467\,10}
F_{589\,11}  
- F_{4679} F_{58\,10\,11} +
F_{4678} F_{59\,10\,11} \nonu \\
&- &  F_{45\,10\,11} F_{6789} + F_{459\,11} F_{678\,10} - F_{459\,10}
F_{678\,11} 
- F_{458\,11} F_{679\,10} + F_{458\,10} F_{679\,11} \nonu \\  
&- &  F_{4589} F_{67\,10\,11} + F_{457\,11} F_{689\,10}-F_{457\,10}
F_{689\,11} + F_{4579} F_{68\,10\,11} - F_{4578} F_{69\,10\,11} \nonu
\\
&- & F_{456\,11} F_{789\,10} + F_{456\,10} F_{789\,11} - F_{4569}
F_{78\,10\,11} + 
F_{4568} F_{79\,10\,11} - F_{4567} F_{89\,10\,11}, \nonu \\
(456) & = & \sin \psi \,
F_{1235} \, F_{789\,10}, \qquad
(457)  =  - \sin \psi \,
F_{1235} \, F_{689\,10},
\nonu \\
(458) & = & \sin \psi \,
F_{1235} \, F_{679\,10}, \qquad
(459)  =  -\sin \psi \,
F_{1235} \, F_{678\,10}, \nonu \\
(45\,10) & = & \sin \psi \,
F_{1235} \, F_{6789}, \qquad
(467)  =  
F_{1235} \, F_{89\,10\,11}, \nonu \\
(468) & = & -
F_{1235} \, F_{79\,10\,11}, \qquad
(469)  =  
F_{1235} \, F_{78\,10\,11}, \nonu \\
(46\,10) & = & F_{1235} ( \cos \psi \, \cot
\mu \, F_{789\,10}- F_{789\,11}), \nonu \\
(46\,11)   & = &  - \cos \psi \, \cot
\mu \,
F_{1235} \, F_{789\,10}, \qquad
(478)  =  
F_{1235} \, F_{69\,10\,11}, \nonu \\
(479)  & = &  -
F_{1235} \, F_{68\,10\,11}, \qquad
(47\,10)  =  
F_{1235} (-\cos \psi \cot \mu \,  F_{689\,10} + F_{689\,11} ), \nonu \\
(47\,11) & = & \cos \psi \, \cot
\mu \,
F_{1235} \, F_{689\,10}, \qquad
(489)  =  
F_{1235} \, F_{67\,10\,11}, \nonu \\
(48\,10) & = & 
F_{1235} (\cos \psi \cot \mu  F_{679\,10} - F_{679\,11}), \qquad
(48\,11)  =  -\cos \psi \, \cot
\mu \,
F_{1235} \, F_{679\,10}, \nonu \\
(49\,10) & = & 
F_{1235} ( -\cos \psi \cot \mu  F_{678\,10} + F_{678\,11}), \nonu \\
(49\,11) & = & \cos \psi \, \cot
\mu \,
F_{1235} \, F_{678\,10}, \qquad
(4\,10\,11)  =  - \cos \psi \,
\cot \mu \,
F_{1235} \, F_{6789}, 
\nonu \\
(567) & = & \sin \psi 
( - F_{1235}  F_{489\,10} + F_{1234}  F_{589\,10}) 
 + 
\frac{
 \cos \mu \, \cos \psi}{\sqrt{1-\cos^2 \psi \, \sin^2 \mu}} 
\, F_{1234} \, F_{89\,10\,11}, \nonu \\
(568) & = & \sin \psi 
( F_{1235}  F_{479\,10} - F_{1234}  F_{579\,10}) 
 - 
\frac{
 \cos \mu \, \cos \psi}{\sqrt{1-\cos^2 \psi \, \sin^2 \mu}} \, 
F_{1234} \, F_{79\,10\,11}, \nonu \\
%
(569) & = & \sin \psi 
( - F_{1235}  F_{478\,10} + F_{1234}  F_{578\,10}) 
 + 
\frac{
 \cos \mu \, \cos \psi}{\sqrt{1-\cos^2 \psi \, \sin^2 \mu}} \,
F_{1234} \, F_{78\,10\,11}, \nonu \\
(56\,10) & = & 
 -   \frac{\cos \mu \, \cos \psi}{ \sqrt{1-\cos^2 \psi \,
  \sin^2 \mu}}
\, F_{1234} \, F_{789\,11} + \csc \mu \,
\sqrt{1-\cos^2 \psi \, \sin^2 \mu}   \, F_{1234} \, F_{789\,10} \nonu
\\
&+ &    \sin \psi \, (F_{1235}\, F_{4789} -F_{1234}\, F_{5789}),  
\nonu \\
(56\,11) & = & -
\, \csc \mu \, \sqrt{1-\cos^2 \psi  \sin^2 \mu} \,  
F_{1234} \, F_{789\,10}, 
\nonu \\
(578) & = & \sin \psi 
(-F_{1235}  F_{469\,10} + F_{1234}  F_{569\,10}) 
 +  \frac{
 \cos \mu \, \cos \psi}{\sqrt{1-\cos^2 \psi \, \sin^2 \mu}} \,
F_{1234}  \, F_{69\,10\,11},
\nonu \\
(579) & = & \sin \psi 
(F_{1235}  F_{468\,10} - F_{1234}  F_{568\,10}) 
 -  \frac{
 \cos \mu \, \cos \psi}{\sqrt{1-\cos^2 \psi \, \sin^2 \mu}}\, 
F_{1234}  \, F_{68\,10\,11},
\nonu \\
(57\,10) & = & 
   \frac{\cos \mu \, \cos \psi}{ \sqrt{1-\cos^2 \psi \,
  \sin^2 \mu}}
\, F_{1234} \, F_{689\,11} - \csc \mu \,
\sqrt{1-\cos^2 \psi \, \sin^2 \mu}   \, F_{1234} \, F_{689\,10} \nonu
\\
&+ &    \sin \psi \, (-F_{1235}\, F_{4689} +F_{1234}\, F_{5689}),  
\nonu \\
(57\,11) & = & 
\, \csc \mu \, \sqrt{1-\cos^2 \psi  \sin^2 \mu} \,  
F_{1234} \, F_{689\,10}, 
\nonu \\
(589) & = & \sin \psi 
(-F_{1235}  F_{467\,10} + F_{1234}  F_{567\,10}) 
 +  \frac{
 \cos \mu \, \cos \psi}{\sqrt{1-\cos^2 \psi \, \sin^2 \mu}} 
\, F_{1234} \,  F_{67\,10\,11},
\nonu \\
(58\,10) & = & 
   -\frac{\cos \mu \, \cos \psi}{ \sqrt{1-\cos^2 \psi \,
  \sin^2 \mu}}
\, F_{1234} \, F_{679\,11} + \csc \mu \,
\sqrt{1-\cos^2 \psi \, \sin^2 \mu}   \, F_{1234} \, F_{679\,10} \nonu
\\
&+ &    \sin \psi \, (F_{1235}\, F_{4679} -F_{1234}\, F_{5679}),  
\nonu \\
(58\,11) & = & -
\, \csc \mu \, \sqrt{1-\cos^2 \psi  \sin^2 \mu} \,  
F_{1234} \, F_{679\,10}, 
\nonu \\
(59\,10) & = & 
  \frac{\cos \mu \, \cos \psi}{ \sqrt{1-\cos^2 \psi \,
  \sin^2 \mu}}
\, F_{1234} \, F_{678\,11} - \csc \mu \,
\sqrt{1-\cos^2 \psi \, \sin^2 \mu}   \, F_{1234} \, F_{678\,10} \nonu
\\
&+ &    \sin \psi \, (-F_{1235}\, F_{4678} +F_{1234}\, F_{5678}),  
\nonu \\
(59\,11) & = & 
\, \csc \mu \, \sqrt{1-\cos^2 \psi  \sin^2 \mu} \,  
F_{1234} \, F_{678\,10}, 
\nonu \\
(5\,10\,11) & = & -
\, \csc \mu \, \sqrt{1-\cos^2 \psi  \sin^2 \mu} \,  
F_{1234} \, F_{6789}, 
\nonu \\
(678)  & = &  -(F_{1235} \,
F_{49\,10\,11} -
F_{1234} \, F_{59\,10\,11}),
\qquad
(679)  =  (F_{1235} \,
F_{48\,10\,11} -
F_{1234} \, F_{58\,10\,11}), 
\nonu \\
(67\,10) & = & 
-F_{1235} \, F_{489\,11} + \cos \psi \, \cot \mu (F_{1235} \,
F_{489\,10} - F_{1234} \, F_{589\,10}) 
\nonu \\
& + & F_{1234} \, F_{589\,11} +  \frac{
\, \csc \mu \, \sin \psi}{\sqrt{1-\cos^2 \psi \,  \sin^2 \mu}} \,
F_{1234} \, F_{89\,10\,11},
\nonu \\
(67\,11) & = & \cos \psi  \cot
\mu 
(-F_{1235}  F_{489\,10} + F_{1234}  F_{589\,10}) 
 -  \frac{
 \csc \mu \, \sin \psi}{\sqrt{1-\cos^2 \psi \, \sin^2 \mu}} 
\, F_{1234} \,  F_{89\,10\,11},
\nonu \\
(689) & = & -(F_{1235} \,
F_{47\,10\,11} -
F_{1234} \, F_{57\,10\,11}), 
\nonu \\
%
(68\,10) & = & 
F_{1235} \, F_{479\,11} - \cos \psi \, \cot \mu (F_{1235} \,
F_{479\,10} - F_{1234} \, F_{579\,10}) 
\nonu \\
& - & F_{1234} \, F_{579\,11} -  \frac{
\, \csc \mu \, \sin \psi}{\sqrt{1-\cos^2 \psi \, \sin^2 \mu}} \,
F_{1234} \, F_{79\,10\,11},
\nonu \\
(68\,11) & = & \cos \psi  \cot
\mu 
(F_{1235}  F_{479\,10} - F_{1234}  F_{579\,10}) 
+  \frac{
 \csc \mu \, \sin \psi}{\sqrt{1-\cos^2 \psi \, \sin^2 \mu}} 
\, F_{1234} \,  F_{79\,10\,11},
\nonu \\
(69\,10) & = & 
-F_{1235} \, F_{478\,11} + \cos \psi \, \cot \mu (F_{1235} \,
F_{478\,10} - F_{1234} \, F_{578\,10}) 
\nonu \\
& + & F_{1234} \, F_{578\,11} +  \frac{
\, \csc \mu \, \sin \psi}{\sqrt{1-\cos^2 \psi \, \sin^2 \mu}} \,
F_{1234} \, F_{78\,10\,11},
\nonu \\
(69\,11) & = & \cos \psi  \cot
\mu 
(-F_{1235}  F_{478\,10} + F_{1234}  F_{578\,10}) 
 -  \frac{
 \csc \mu \, \sin \psi}{\sqrt{1-\cos^2 \psi \, \sin^2 \mu}} 
\, F_{1234}  \, F_{78\,10\,11},
\nonu \\
(6\,10\,11) & = & \cos \psi  \cot
\mu 
(F_{1235}  F_{4789} - F_{1234}  F_{5789}) 
 +  \frac{
 \csc \mu \, \sin \psi}{\sqrt{1-\cos^2 \psi \, \sin^2 \mu}} 
\, F_{1234} \,  F_{789\,11},
\nonu \\
(789) & = & (F_{1235} \,
F_{46\,10\,11} -
F_{1234} \, F_{56\,10\,11}), 
\nonu \\
(78\,10) & = & 
-F_{1235} \, F_{469\,11} + \cos \psi \, \cot \mu (F_{1235} \,
F_{469\,10} - F_{1234} \, F_{569\,10}) 
\nonu \\
& + & F_{1234} \, F_{569\,11} +  \frac{
\, \csc \mu \, \sin \psi}{\sqrt{1-\cos^2 \psi \,  \sin^2 \mu}} \,
F_{1234} \, F_{69\,10\,11},
\nonu \\
(78\,11) & = & \cos \psi  \cot
\mu 
(-F_{1235}  F_{469\,10} + F_{1234}  F_{569\,10}) 
 -  \frac{
 \csc \mu \, \sin \psi}{\sqrt{1-\cos^2 \psi \, \sin^2 \mu}} 
\, F_{1234} \, F_{69\,10\,11},
\nonu \\
(79\,10) & = & 
F_{1235} \, F_{468\,11} - \cos \psi \, \cot \mu (F_{1235} \,
F_{468\,10} - F_{1234} \, F_{568\,10}) 
\nonu \\
& - & F_{1234} \, F_{568\,11} -  \frac{
\, \csc \mu \, \sin \psi}{\sqrt{1-\cos^2 \psi \, \sin^2 \mu}} \,
F_{1234} \, F_{68\,10\,11},
\nonu \\
(79\,11) & = & \cos \psi  \cot
\mu 
(F_{1235}  F_{468\,10} - F_{1234}  F_{568\,10}) 
 + \frac{
 \csc \mu \, \sin \psi}{\sqrt{1-\cos^2 \psi\,  \sin^2 \mu}} 
\, F_{1234}\,  F_{68\,10\,11},
\nonu \\
(7\,10\,11) & = & \cos \psi  \cot
\mu 
(-F_{1235}  F_{4689} + F_{1234}  F_{5689}) 
 -  \frac{
 \csc \mu  \, \sin \psi}{\sqrt{1-\cos^2 \psi \, \sin^2 \mu}} 
\, F_{1234} \, F_{689\,11},
\nonu \\
(89\,10) & = & 
-F_{1235} \, F_{467\,11} + \cos \psi \, \cot \mu (F_{1235} \,
F_{467\,10} - F_{1234} \, F_{567\,10}) 
\nonu \\
& + & F_{1234} \, F_{567\,11} +  \frac{
\, \csc \mu \, \sin \psi}{\sqrt{1-\cos^2 \psi \, \sin^2 \mu}} \,
F_{1234} \, F_{67\,10\,11},
\nonu \\
(89\,11) & = & \cos \psi  \cot
\mu 
(-F_{1235}  F_{467\,10} + F_{1234}  F_{567\,10}) 
 -  \frac{
 \csc \mu \, \sin \psi}{\sqrt{1-\cos^2 \psi \, \sin^2 \mu}} 
\, F_{1234} \,  F_{67\,10\,11},
\nonu \\
(8\,10\,11) & = & \cos \psi  \cot
\mu 
(F_{1235}  F_{4679} - F_{1234}  F_{5679}) 
 +  \frac{
 \csc \mu \, \sin \psi}{\sqrt{1-\cos^2 \psi \, \sin^2 \mu}} 
\,F_{1234}\,  F_{679\,11},
\nonu \\
(9\,10\,11) & = & \cos \psi  \cot
\mu 
(-F_{1235}  F_{4678} + F_{1234}  F_{5678}) 
 -  \frac{
 \csc \mu \, \sin \psi}{\sqrt{1-\cos^2 \psi \, \sin^2 \mu}} 
\, F_{1234} \,  F_{678\,11}.
\nonu
\eea
Note that there are nonzero components
$(45m)$ where $m=6, \cdots, 10$ while these are vanishing for $G_2$-invariant flow.


\end{document}